\newcommand{\blind}{1}
\documentclass[12pt]{article}
\usepackage{amsmath,epsfig,amssymb,amsfonts,amsthm,epsfig,verbatim,enumitem,bm,mathrsfs,xr}
\usepackage[colorlinks=true,linkcolor=blue,citecolor=blue,urlcolor=blue]{hyperref}
\usepackage{natbib,graphicx,algorithm,algorithmic}
\usepackage{color,graphicx,lscape,longtable,mathabx, float, multirow}
\usepackage{hyperref,subfigure, setspace,caption,diagbox}
\usepackage{natbib}
\usepackage{marginnote}
\usepackage{soul}
\usepackage{ulem} 

\newcommand{\Rmnum}[1]{\uppercase\expandafter{\romannumeral #1\relax}}

\renewcommand{\d}{\stackrel{d}{\rightarrow}}

\graphicspath{{./figure/}}

\newtheorem{rem}{Remark}

\newtheorem{Ex}{Example}

\def\bse{\begin{eqnarray*}}
	\def\ese{\end{eqnarray*}}
\def\be{\begin{eqnarray}}
\def\ee{\end{eqnarray}}
\def\bsq{\begin{equation*}}
\def\esq{\end{equation*}}
\def\bq{\begin{equation}}
\def\eq{\end{equation}}

\def\var{\hbox{var}}

\def\cov{\hbox{cov}}

\def\wh{\widehat}
\def\wt{\widetilde}

\def\cov{\mbox{cov}}

\def\diag{\mbox{diag}}

\def\sumi{\sum_{i=1}^n}

\def\sumj{\sum_{j=1}^p}

\def\bTheta{\boldsymbol\Theta}
\def\ba{\boldsymbol\alpha}

\def\bta{\boldsymbol\eta}
\def\bzeta{\boldsymbol\zeta}

\def\bb{{\boldsymbol\beta}}

\def\LLambda{{\bf\Lambda}}
\def\0{{\bf 0}}

\def\R{{\bf R}}
\def\Q{{\bf Q}}

\def\B{{\bf B}}
\def\E{{\bf E}}

\def\K{{\bf K}}

\def\PPhi{{\bm\Phi}}
\def\PPsi{{\bm\Psi}}

\def\h{{\bf h}}
\def\b{{\bf b}}
\def\I{{\bf I}}
\def\M{\mbox{ $\mathcal{M}$}}
\def\M{{\bf M}}

\def\K{{\bf K}}
\def\k{{\bf k}}
\def\t{{\bf t}}

\def\u{{\bf u}}

\def\W{{\bf W}}

\def\w{{\bf w}}
\def\X{{\bf X}}

\def\I{{\bf I}}

\def\bGamma{{\bf \Gamma}}

\def\bSig{{\bf \Sigma}}

\def\RX{\mathbb{R}}
\def\NX{\mathbb{N}}

\def\bq{\begin{equation}}
\def\eq{\end{equation}}

\def\wh{\widehat}
\def\wt{\widetilde}

\def\log{{\rm log}}

\def\squarebox#1{\hbox to #1{\hfill\vbox to #1{\vfill}}}

\def\var{\hbox{var}}
\def\cov{\hbox{cov}}

\def\bse{\begin{eqnarray*}}
	\def\ese{\end{eqnarray*}}
\def\be{\begin{eqnarray}}
\def\ee{\end{eqnarray}}
\def\bsq{\begin{equation*}}
\def\esq{\end{equation*}}
\def\bq{\begin{equation}}
\def\eq{\end{equation}}
\def\Ex{\mathbb{E}}
\def\wh{\widehat}
\def\wt{\widetilde}

\def\cov{\mbox{cov}}
\def\boxit#1{\vbox{\hrule\hbox{\vrule\kern6pt\vbox{\kern6pt#1\kern6pt}\kern6pt\vrule}\hrule}}

\usepackage{amsmath,amssymb}
\DeclareSymbolFont{EulerExtension}{U}{euex}{m}{n}
\DeclareMathSymbol{\euintop}{\mathop} {EulerExtension}{"52}
\DeclareMathSymbol{\euointop}{\mathop} {EulerExtension}{"48}

\renewcommand{\baselinestretch}{2}

\begin{document}
\def\spacingset#1{\renewcommand{\baselinestretch}%
	{#1}\small\normalsize} \spacingset{1}

\if1\blind{	
	\title{\bf Factor-guided functional PCA \\ for high-dimensional functional data}
	 \author{Shoudao Wen and Huazhen Lin\thanks{
	  		Corresponding author. Email: linhz@swufe.edu.cn. The research were partially supported by National Natural
	  		Science Foundation of China (Nos. 11931014 and 11829101).}\hspace{.2cm}\\
	  		Center of Statistical Research and  School of Statistics,\\ Southwestern University of
	  		Finance and Economics}	
	 \maketitle
}\fi	
	
\if0\blind{	
	  \bigskip
	  \bigskip
	  \bigskip
	\begin{center}
	{\LARGE\bf Factor-guided functional PCA \\ for high-dimensional functional data}		
	\end{center}
	\medskip
}\fi

\bigskip	
\begin{abstract}
The literature on high-dimensional functional data  focuses on either the dependence over time or  the correlation among functional variables. In this paper, we propose a  factor-guided functional principal component analysis (FaFPCA)   method to  consider  both temporal  dependence  and   correlation of variables  so that the extracted features are as sufficient as possible. In particular, we use a factor process to consider the correlation among high-dimensional functional variables and then apply functional principal component analysis (FPCA) to  the factor processes to address the dependence over time. Furthermore, to  solve the computational problem arising from triple-infinite dimensions, we creatively build some moment equations to estimate loading, scores and eigenfunctions in closed form without rotation.
Theoretically, we  establish  the asymptotical  properties of  the proposed estimator. 
Extensive simulation studies demonstrate that our proposed method outperforms other competitors in terms of accuracy and computational cost.   The proposed method is  applied to analyze the  Alzheimer's Disease Neuroimaging Initiative (ADNI) dataset, resulting in
higher prediction accuracy and 41 important ROIs that are associated with Alzheimer's disease, 23 of which have been confirmed by the literature.
\end{abstract}

\vfill	

\newpage
\spacingset{1.9}
\addtolength{\textheight}{.5in}	
\section{Introduction}
\label{sec1}	
Advances in technology have enabled us to collect an increasing amount of high-dimensional 
	functional-type data
	in areas such as medicine, economics, finance, traffic and climatology. 
	Some of these datasets include the
Wharton Research Data Services (WRDS) dataset; the Surveillance, Epidemiology, and End Results (SEER) dataset; the Alzheimer's Disease Neuroimaging Initiative (ADNI) dataset \citep{mueller2005alzheimer} and so on. 
Since the seminal work of \cite{ramsay:1982},  great attention 
has been given to functional data analysis (FDA);
for example, the monographs by
\cite{ramsay.silverman:2002,ramsay.silverman:2005,ferraty.vieu:2006,horvath2012inference,hsing.eubank:2015,kokoszka.reimherr:2017}. 
Due to the infinite dimensions of functional data, the compression and dimension reduction of functional data  are crucial, 
in which  a popular and powerful tool is
functional principal component analysis (FPCA).    
FPCA  takes advantage of the dependence over time and allows us to compress 
information without much loss.
The conventional FPCA approach, termed uFPCA,  was proposed for analyzing univariate  functional data  
\citep{rice1991estimating,james2000principal,yao.muller.wang:jasa2005,hall2008modelling,li2010deciding, zhou2018efficient, zhong2021cluster}.
When multivariate functional data of $p$-dimension is available,
\cite{panaretos2010second}, \cite{lian2013shrinkage}, \cite{chiou2014linear}, \cite{zhu2016bayesian}, \cite{grith2018functional} and \cite{han2018smooth} applied uFPCA to each of them.
Since  the eigenfunctions and principal component scores of  each curve are estimated independently, the dependence among  the $p$ curves is easily ignored by them. 
Multivariate FPCA (mFPCA) hence
was  introduced  by applying the same scores to  each component of multivariate curves so that  the dependence among the curves could be taken into account \citep{ramsay.silverman:2005, chiou2014multivariate, jacques2014model, gorecki2018selected, happ2018multivariate, happ2019general, wong2019partially}.
	
However, it is  challenging  to extend mFPCA to high-dimensional functional data. First, in mFPCA,  because each curve has its own eigenfunctions or principal components,  intensive computation  is required for high-dimensional functional data. 
Second, mFPCA  expresses  all the  curves by the same score vector, which may be  insufficient to represent features in high-dimensional functional data, and then further analysis based on the obtained scores is also inadequate. Finally, the theory developed for  fixed-dimensional mFPCA is not applicable to high-dimensional functional data.

	
Some studies have considered
high-dimensional functional data, and  can be roughly divided into three categories: the  frameworks of  uFPCA (uFPCA-HD), sparse FPCA (sFPCA-HD) and  factor models (FM-HD).
uFPCA-HD  applies  uFPCA  to each functional variable for  a score matrix with a low-rank structure  \citep{ gao2019high,chang2020autocovariance,guo2020consistency,solea2021nonparametric,Fang2021finite,tang2021forecasting,tang2021multi}.  However, since the extracted scores represent the most important features  of each functional variable,  the information about the correlation among the functional variables is prone to be ignored. \cite{hu2022sparse} proposed sparse FPCA  
by
 assuming that only a small number of
the high-dimensional functional variables have energies. This assumption usually does  not hold
since the  correlation, which is essential for high-dimensional variables, may cause the difficulty  
of identifying functions with zero or small energies.
On the other hand, the FM-HD emphasizes the correlation among $p$ functional variables. Particularly,
\cite{hu2021dynamic} proposed dynamic principal component analysis (DPCA) based on standard factor analysis at each time {point}. 
\cite{tavakoli2021high} and \cite{guo2022factor} directly extended the  factor model from high-dimensional vector to high-dimensional functional data  by  regarding functional data as  elements in Hilbert space. Hence, the dependence structure over time might be largely  ignored by FM-HD.
These concerns are also confirmed in  Figure \ref{model2}, which shows that FM-HD performs better, but  sFPCA-HD and uPFCA-HD worsen when  the functional variables become correlated; however, the opposite phenomenon occurs when the dependence over time 
becomes stronger. 
	
	
In this paper,  we introduce a  factor-guided 
FPCA (FaFPCA) approach to utilize  the dependence both over time and  among  functional variables so that the extracted  features from  high-dimensional functional data 
are sufficient. Specifically, we firstly propose  factor processes to  organize the correlation of  functional variables,  and then apply uFPCA to each component of the factor processes to address the dependence over time.  
Although the idea is quite natural,  it makes remarkable difference with the existing methods.
For example, due to the correlation among functional variables, uFPCA-HD may extract a large number of duplicate features, while  ignore some features associated with 
correlation information among  variables because of   the first applying  FPCA  to each functional variable. 
In contrast, by
firstly utilizing factor processes, the proposed FaFPCA   efficiently organize the correlation information  and  avoid overlapping features. Moreover, with applying  FPCA to the extracted factor processes, FaFPCA imposes a low-rank structure on the loading function of FM-HD,   further captures the information of temporal dependence, and hence is more  efficient than FM-HD for dimension reduction.

However, the proposed strategy  incurs complicated 
computations due to the triple-infinite dimension, i.e., the
$\B$ loadings, the $\bzeta$ scores and  the  $\PPhi(\cdot)$ eigenfunctions are infinite with  
$\PPhi(\cdot)$ containing $K\times q$ functions,  where $K\rightarrow\infty$. 
A possible estimate method  is to iterate among  three unknown terms.  The computation is intensive and more problematic,  the  iteration usually fails to converge without appropriate initial values, which are almost impossible  to obtain in the current case. 
In this paper,  on the basis of  the proposed model and identification conditions, we  creatively build some moment equations upon which  we  estimate $\B$, $\bzeta$ and  $\PPhi(\cdot)$ in closed form without rotation.   Furthermore, the theoretical properties of the proposed estimator, including the convergence rate and asymptotical normality, are established. The theoretical results suggest that
the high dimension $p$ is a blessing of dimensionality instead of a  curse,
and
both the controlled complexity of  functional data and the limited number of eigenfunctions can significantly improve the estimation accuracy.
These findings highlight  the importance of using  the  dependence both over time and  among  functional variables.
Finally, we apply the proposed method to analyze the dataset from ADNI %
and obtain a higher prediction accuracy than the state-of-the-art methods for high dimensional functional data,
such as uFPCA-HD,  sFPCA-HD and FM-HD  \citep{gao2019high,hu2021dynamic, hu2022sparse, tavakoli2021high},  see Figure \ref{npex_realdata}. Moreover, based on the proposed FaFPCA approach, we identify the atrophy of  41 out of 90 ROIs  that are significantly associated with Alzheimer’s disease, where 23  ROIs  have been verified by the literature. 

	The rest of the paper is organized as follows. In Section 2, we
	introduce the 
	FaFPCA and the estimate method.  In Section 3, we establish the main theoretical results. The performance of the proposed estimation procedure is evaluated through simulation studies and real data analyses in Sections 4 and 5, respectively. A brief discussion about further research is given in Section 6. All proofs are 
	detailed in
	the Supplementary Materials section. The FaFPCA R package based on our method is available on our GitHub home page. 

\section{Model and Estimation}
	
\subsection{FaFPCA Model}
	\label{model}
	
Suppose that the observations $\X_i(t), i = 1, \cdots, n$ are independent identically
	distributed (iid), with $\X_i(t)= (X_{i1}(t), \cdots,
	X_{ip}(t))'$ being the high-dimensional functional variables  for  $p\gg n$.
		For simplicity, we assume that the mean of $X_{ij}(t)$  has already
	been removed; namely, $\Ex\big(X_{ij}(t)\big)=0$ for any $j$ and $t\in [0,\ 1]$. To  reflect the situation of irregular and possibly
	subject-specific time points, we
	assume that $\X_{i}(\cdot)$ are measured at
	$\t_i=(t_{i1},\cdots, t_{in_i})^\prime$.
	
	Illustrated by the idea of the factor analysis   \citep{bai2002determining,bai2003inferential,bai2006confidence,bai2013principal,bai2013statistical}, we assume the $\X_i(t)$
	are correlated because 
	they share a  vector of factor processes $\h_i(t)$.
	The following model is considered:
	\begin{eqnarray}\label{eq:bh}
	\label{eq:model10}
	\X_i(t) = \B \h_i(t) + \u_i(t), 
	\end{eqnarray}
	where $\h_i(t) = (h_{i1}(t),\cdots,
	h_{iq}(t))'$ is a $q$-dimensional vector of factor processes with $q\ll p$, $\B = (\b_1,\cdots, \b_p)'=(b_{jk})_{p \times q}$ is a  deterministic matrix,  and $\u_i(t)=(u_{i1}(t),\cdots,u_{ip}(t))'$ denotes  measurement error independent of $\h_i(\cdot)$ with $\Ex(u_{ij}(t)) = 0$ and $\var(u_{ij}(t))=\sigma^2$. Furthermore, to consider the dependence of functional data over time,  we apply the Kauhunen-Lo$\rm\grave{e}$ve expansion \citep{Ash1975Topics} 
	to the factor process $h_{ij}(t)$,  and have,
	\begin{eqnarray}
	\label{eq:kl1}
	h_{ij}(t)=\sum_{k=1}^{\infty}\xi_{ijk}\phi_{jk}(t), 
	\end{eqnarray}
	where $\phi_{jk}(t)$ is the $k$th orthonormal eigenfunction of the covariance function $C_j(s,t)=\cov(h_{ij}(s),h_{ij}(t))$ for factor process $j$, which satisfies $\int\phi_{jk}(t)\phi_{jk'}(t)dt =1$ if $k=k'$, and it is $0$ otherwise; $\xi_{ijk},k=1,2,\cdots$ are
	the functional principal component scores for the stochastic process
	$h_{ij}(t)$ with $\Ex(\xi_{ijk})=0$, $\var(\xi_{ijk})=\rho_{jk}$ 
	and
	$\mathrm{cov}(\xi_{ijk},\xi_{ijk'})=0$ if $k\neq k'$.  $\rho_{jk}$ is the
	eigenvalue corresponding to the eigenfunction $\phi_{jk}(\cdot)$,   with the constraint of $\rho_{j1}\geq\rho_{j2}\geq\cdots>0$  and $\sum_{k=1}^\infty \rho_{jk}<\infty$ for any $j=1,\cdots, q$, which implies that
 $\sup_{t\in[0,1]}\Ex\big( \sum_{k=1}^{K}\xi_{ijk}\phi_{jk}(t)-\sum_{k=1}^{\infty}\xi_{ijk}\phi_{jk}(t)\big)^2\to 0$ as $K\rightarrow \infty$, we  hence suppose that
	\begin{eqnarray}
		\label{eq:kl}
		h_{ij}(t)=\sum_{k=1}^{K}\xi_{ijk}\phi_{jk}(t). 
	\end{eqnarray} Model (\ref{eq:kl}) has been extensively studied in the literature of 
	FPCA for the case when $K$ is fixed
	\citep{james2000principal,yao.muller.wang:jasa2005,hall2008modelling, zhou2018efficient}. To improve  flexibility,  we allow  $K\rightarrow \infty$  
	\citep{hall2006properties,kong2016partially,lin2018mixture}. In addition, $K$ can vary with $j$, and we use the same  $K$  for  $j=1,\cdots, q$  for simple notation.

	Let $\bzeta_{ij}=(\xi_{ij1},\cdots,\xi_{ijK})'$, $\bzeta_{i}=(\bzeta'_{i1},\cdots,\bzeta'_{iq})'$ and $\PPhi(t)=\diag(\PPhi_1(t),\cdots,\PPhi_{q}(t))$, where $\PPhi(t)$ 
	is a $Kq\times q$ block diagonal matrix with block $j$ being $\PPhi_j(t)=(\phi_{j1}(t),\cdots,\phi_{jK}(t))'$. 
	Combined with the expression in (\ref{eq:kl}), the model in  (\ref{eq:model10}) can be written as the following factor-guided functional principal component analysis (FaFPCA)  model:
	\begin{eqnarray}
	\label{eq:model1}
	\X_i(t)=\B\PPhi'(t)\bzeta_i+\u_i(t).
	\end{eqnarray}
As mentioned above, model (\ref{eq:model10}) highlights  the correlation  among the functional variables, and the model (\ref{eq:kl}) expresses the temporal dependence of the functional data. As a result, the proposed FaFPCA  model in (\ref{eq:model1}) simultaneously  
organizes the two kinds of dependence structures  so that a simple and sufficient expression for the  high-dimensional functional  may be  obtained.
Different from the  framework of  uFPCA (uFPCA-HD),  
the proposed FaFPCA  utilizes  factor processes to extract features from functional variables, hence can  efficiently  avoid overlapping features and  organize the correlation information among functional variables. In addition,  compared with  the framework of factor models (FM-HD),  
FaFPCA captures the information of temporal dependence by  imposing  a low-rank  decomposition $\B\PPhi'(t)$ on the  loading function in FM-HD, and hence  is more structured and  efficient than FM-HD for dimension reduction. 

Obviously, model (\ref{eq:model1}) is not identifiable. We impose three
	constraints for identifiability:
	\begin{description}
		\item[(I1)]	$p^{-1}\B'\B=\I_{q}$ and the first nonzero element of each column of $\B$ is positive.
		\item[(I2)] $\bzeta'\bzeta$ is diagonal with decreasing diagonal entries, where
		$\bzeta=(\bzeta_1,\cdots,\bzeta_n)'$. 
		\item[(I3)] $\int\PPhi(t)\PPhi(t)' dt=\I_{Kq}$ and $\phi_{jk}(0)>0$.
	\end{description}
Conditions (I1) and (I2), which are restrictions  on loadings and factors, are commonly used in the literature of factor analysis \citep{bai2013principal, fan2013large,li2018embracing,asz010,liu2021generalized}. 
Condition (I3), which restricts the eigenfunctions, is  widely used in the literature of  FPCA   \citep{james2000principal,yao.muller.wang:jasa2005,hall2008modelling, zhou2018efficient}.
We state the identifiability in the following proposition, and its proof is 
detailed in the Supplementary Materials B.
	
	\newtheorem{Proposition1}{Proposition}
	\begin{Proposition1}
		Under identification conditions (I1)-(I3) and (A1)-(A4) in Section \ref{sec:theo}, $\B, \bzeta$ and $\PPhi(t)$ are unique as
		$n\to \infty$ and $p\to \infty$ for both finite $K$ and divergent $K$.
	\end{Proposition1}

\subsection {Estimation Procedure}
	\label{est11}

	The estimation involves three terms $\B$, $\bzeta$ and $\PPhi(\cdot)$. Although model  (\ref{eq:model1})
	is a kind of factor model,  the existing  methods for factor analysis
	are not feasible for the estimation of model  (\ref{eq:model1}) because  
	the  matrix-structured data required {for factor analysis
	is not available for functional data due to their individual-specific observation times.    A possible method  is to iterate among the three terms $\B$, $\bzeta$ and $\PPhi(\cdot)$, which 
	is considerably computationally intensive
	because 
	all of them are infinite. 
	Moreover,  the iteration usually fails to converge without appropriate initial values, which are  almost impossible to  obtaine  in the current situation. 
Here, we avoid the iteration method and solve the computational problem  by  building some moment equations. Concretely,  we denote the covariance matrices of $\X_i(t)$ and $\u_i(t)$ by
$\bSig_{\X}(t)$ and $\bSig_{\u}(t)$, respectively. Let
	$\wt\bSig_{\X}=\int\bSig_{\X}(t) dt$, $\wt\bSig_{\u}=\int\bSig_{\u}(t) dt$ and  $\LLambda_{\bzeta}=\diag\big(\sum_{k=1}^{K} \var(\xi_{i1k}),\cdots,\sum_{k=1}^{K} \var(\xi_{iqk})\big)$.
	By calculating the covariance matrix and 
	integrating over $t$ on both sides of (\ref{eq:model1}), we have
	\begin{eqnarray}\label{eq:wtSx}
	\label{eq:B1}
	\wt\bSig_{\X}
	=\B\LLambda_{\bzeta}\B'+\wt\bSig_{\u},
	\end{eqnarray}
	which implies  that $p^{-1/2}\B$ is the eigenvector of $\wt\bSig_{\X}$. Considering the identification condition $p^{-1}\B'\B=\I_{q}$, we estimate $p^{-1/2}\wh\B$ for $p^{-1/2}\B$ by  the  orthogonal eigenvectors of the numerical approximation version of $\wt\bSig_{\X}$; i.e.,
	\begin{eqnarray}
	\label{eq:B11}
	\wh\B=p^{1/2}\E_{eigen}\Big(n^{-1}\sumi n^{-1}_i\sum _{l=1}^{n_i}\X_i(t_{il})\X'_i(t_{il});q\Big),
	\end{eqnarray}
	where $\E_{eigen}(A; q)$ is a matrix  composed of   the orthogonal eigenvectors corresponding to the $q$ largest eigenvalues of matrix $A$.

	Now we consider $\PPhi$ and $\bzeta$.
	Without loss of generality, we assume that the support of $t_{ij}$  is $[0, \ 1]$ by  proper scaling.  $\M(\cdot)=(M_1(\cdot), \ \cdots, \ M_{\tau_n}(\cdot))'$ denote a
	vector of B-spline basis functions  on $[0, \ 1]$ with $\tau_n=O(n^{v})$; then we have $\phi_{jk}(t)\approx\bTheta'_{jk}\M(t)$. Let $\bTheta_j=(\bTheta_{j1},\cdots,\bTheta_{jK})'\in\RX^{K\times \tau_n}.$
	With the  B-spline approximation, identification condition (I2) on $\PPhi(\cdot)$ can be written as  $\tau^{-1}_n\bTheta_{j}\bTheta'_{j}=\I_{K}, j=1,\cdots, q$  and
	$\tau_n\int \M(t)\M'(t) dt=\I_{\tau_n}$.
	Then by (\ref{eq:kl}),
	we have
	\begin{eqnarray}
	\label{eq:h1}
	h_{ik}(t_{il})= \PPhi'_k(t_{il})\bzeta_{ik}\approx \M'(t_{il})\bTheta'_k\bzeta_{ik}.
	\end{eqnarray}
	On the other hand, by multiplying $p^{-1}\B'$ on both sides of (\ref{eq:model10}), we have $
	p^{-1}\B'\X_i(t) \approx p^{-1}\B' \B \h_i(t).$  
	Since  $p^{-1}\B'\B=\I_{q}$, we obtain
	\begin{eqnarray}
	\label{eq:h2}
	h_{ik}(t_{il})\approx p^{-1}\sumj b_{jk}X_{ij}(t_{il}).
	\end{eqnarray}
	Combining (\ref{eq:h1}) and (\ref{eq:h2}), we obtain,
	\begin{eqnarray}\label{eq:h}
	p^{-1}\sumj b_{jk}X_{ij}(t_{il}) \approx  \M'(t_{il})\bTheta'_k\bzeta_{ik}.
	\end{eqnarray}	
Then, multiplyling $\M(t_{il})$ on both sides of (\ref{eq:h}) and making  the summation over the observation times of individual $i$,  we have
	$$
	\big\{\sum_{l=1}^{n_i}\M(t_{il})\M'(t_{il})\big\}^{-1}\big\{\sum_{l=1}^{n_i}\sumj p^{-1}\M(t_{il})b_{jk}X_{ij}(t_{il})\big\}\approx\bTheta'_k\bzeta_{ik},
	$$
which is a factor model with response $\w_{ik}=\big\{\sum_{l=1}^{n_i}\M(t_{il})\M'(t_{il})\big\}^{-1}\big\{\sum_{l=1}^{n_i}\sumj p^{-1}\M(t_{il})\wh b_{jk}X_{ij}(t_{il})\big\}$, factor  $\bzeta_{ik}$ and loading $\bTheta_k$, where
$\hat b_{jk}$ is  the estimator from (\ref{eq:B11}).
Here $\W_k=(\w_{1k}, \cdots, \w_{nk})'\in\RX^{n\times\tau_n}$ and $\bzeta_{[k]}=(\bzeta_{1k}, \cdots, \bzeta_{nk})'\in\RX^{n\times K}$; hence, we estimate $(\bzeta_{[k]},\bTheta_k)$  by
\begin{eqnarray}
\label{zt}
	(\wh\bzeta_{[k]},\wh\bTheta_k)=\mathop{\arg\min}_{\bzeta_{[k]},
		\bTheta_k}\Big\|\W_k-\bzeta_{[k]} \bTheta_k\Big\|^2_F,
\end{eqnarray}
with $\|\cdot\|_F$ being the Frobenius-norm of a matrix. 	Following \cite{bai2013principal}, which states that  the  factors can be estimated asymptotically  by  principal component analysis (PCA),  we obtain estimators for $\bTheta_k$ and $\bzeta_{[k]}$, respectively by
	\begin{eqnarray}
	\label{eq:theta}
	&&\wh{\bTheta}_{k}'=\E_{eigen}(\W_k'\W_k; K),\\
	&&\wh\bzeta_{[k]}=\tau_n^{-1/2}\W_k\times\E_{eigen}(\W_k'\W_k; K),
	\label{eq:zeta}
	\end{eqnarray}
	for $k=1,\cdots, q$. Finally,  we estimate $\phi_{jk}(t)$ by $\wh\phi_{jk}(t)=\wh\bTheta'_{jk}\M(t)$.

	
	The estimators (\ref{eq:B11}), (\ref{eq:theta}) and (\ref{eq:zeta}) for $\B$, $\bTheta$ and $\bzeta$, respectively, have closed forms based on eigenvalue decomposition without any rotation,  and hence, the computation and programming are very simple. 

\subsection{Determining the number of factors  and eigenfunctions}
\label{section:2.3}
We  need to  determine the dimension of the factor components $q$ and the number of eigenfunctions $K$. Since we transform the estimation for $\B$, $\bzeta$ and $\PPhi(\cdot)$  to
	the problem of principal component analysis, 
	$q$ and $K$  can be selected by calculating the proportion of variability explained by  the principal components \citep{james2000principal,happ2018multivariate}.
	Particularly, we choose $q$ and $K$ such that $$\sum _{j=1}^{q}\lambda_j\left(\sum _{i=1}^nn^{-1}_i\sum_{l=1}^{n_i} \X_i(t_{il})\X_i'(t_{il})\right)/\sumj\lambda_j\left(\sum _{i=1}^nn^{-1}_i\sum_{l=1}^{n_i} \X_i(t_{il})\X_i'(t_{il})\right)>95\%,$$
	$$
	\min_{k\in\{1,\cdots,q\}}\sum _{j=1}^{K}\lambda_j(\W'_k\W_k)/\sum _{j=1}^{\tau_n}\lambda_j(\W'_k\W_k)>95\%,
	$$
	respectively,  where  $\lambda_k(A)$ is the $k$th eigenvalue of $A$.
	In practice, we can choose a different $K_{k}$ for $k=1,\cdots,q$  so that $
	\sum_{j=1}^{K_{k}}\lambda_j(\W'_k\W_k)/\sum _{j=1}^{\tau_n}\lambda_j(\W'_k\W_k)>95\%
	$ is more flexible. 

\section{Theoretical Properties}
	\label{sec:theo}
	Let $K=O(n^{e})$, the following assumptions are required to establish the theoretical properties of $\wh\B$, $\wh\bzeta$ and $\wh\PPhi(\cdot)$.
	\begin{itemize}	
		\item[(A1)] As $n\to\infty$,  $\|n^{-1}\bzeta\bzeta'-\bSig_{\bzeta}\|_2\to 0$ and 
		$\bSig_{\bzeta}=\diag(\bSig_{\bzeta,1},\cdots,\bSig_{\bzeta,q})$
		is a positive definite diagonal matrix, where $\bSig_{\bzeta,j}=\diag(\bSig_{\bzeta,j1},\cdots,\bSig_{\bzeta,jK})$. There exist two positive constants
		$C_1,C_2$ such that  $C_1\leq\sum_{k=1}^{K}\bSig_{\bzeta,jk}\leq C_2$ for each $j=1,\cdots,q$.	
		\item[(A2)] There exist  positive constants $C$, $a_1, a_2$ and $C_1,C_2$, such that  (1)
		sup$_j\|\b_j\|_2\leq C$; (2)  for any $s>0$, $P(\sup_{j,k}|\xi_{ijk}|>s)\leq\exp(-(s/C_1)^{a_1})$ and $P(\sup_{j}|u_{ij}(t)|>s)\leq\exp(-(s/C_2)^{a_2})$.
		
		\item[(A3)] The random errors $\u_i(t)$ are independent of  each other and  $\bzeta_i$, and	$\Ex(u_{ij}(t))=0$,
		$\sum_{j'=1}^p \left|\Ex[u_{ij}(t)u_{ij'}(t)]\right|\leq C$ for each $j$ and uniformly over $t$.
		Furthermore, there exists $\delta\geq 4$ such that
		$\Ex\big|p^{-1/2}\sumj \big[u^2_{ij}(t)-\Ex(u^2_{ij}(t))  \big]\big|^{\delta}\leq C$ and
		$\Ex\big\|p^{-1/2}\sumj \b_ju_{ij}(t)\big\|^{\delta}_2\leq C$  uniformly over $t$.
		\item[(A4)]  $p^{-1/2}\sumj\b_ju_{ij}(t)\d N(0,\bGamma(t))$ as $p \to \infty$, where $\bGamma(t)=\lim_{p\to\infty}\frac{1}{p}\sum_{j,j'=1}^p \b_j\b_{j'}' \Ex(u_{ij}(t)u_{ij'}(t))$.
		\item[(A5)] 
		Denote that $w_j$ is the $j$-th knot for $\M(\cdot)$,
		$\triangle_1=\mathop{\max} _{j}|w_j-w_{j-1}|$  and  $\triangle_2=\mathop{\min} _{j}|w_j-w_{j-1}|$. We assume $\triangle_1=O(n^{-v})$, where $0<v<1/2$ and $\triangle_1/\triangle_2$ is bounded.
		\item[(A6)] 
		$\omega=k+s$ for $k\in\NX_{+}$ and $s\in(0,\ 1]$, and  	
		$\mathscr{H}_{\omega}=\Big\{g(\cdot):|g^{(k)}(x)-g^{(k)}(y)|\leq C|x-y|^{s}\ \text{for any}\  x, y\Big\}.$
		We suppose the true functions  $\{\phi_{jk0}:j=1,\cdots,q;k=1,\cdots,K\}\in\mathscr{H}_{r}$
		with $r\geq 2$.
	\end{itemize}
	
	Condition (A1) is  commonly used  in the factor model \citep{bai2013principal} and implies the existence of $Kq$ factors and the boundedness of corresponding variances.
	Condition (A2) is similar to Assumption 3.2 in \cite{bai2013statistical} for the linear factor model. Particularly, condition (A2.1) requires the loadings to be uniformly bounded,  and condition (A2.2) is a requirement on factors and random errors having  an exponential tail, which is used  to establish  the uniform convergence rates of $\wh\b_j$ and $\wh\bzeta_i$.
	Condition (A3)  allows  the random errors $u_{ij}(t)$ to be correlated with each other to some extent. 
	Condition (A4) is a technique condition.
	Condition (A5) implies that the spline knots are uniform and are  commonly used for spline approximation theory.
	Condition (A6) is a regular smoothing condition on the  functions.

	\newtheorem{theorem}{Theorem}
	\begin{theorem} \label{th1}
		Under Conditions (A1)-(A6),  we have
		\begin{eqnarray*}
			\hspace{-0.6cm}	\begin{split}
				\|\wh\b_j-\b_{j0}\|_2=&O_p(\mathcal{R}_{pn}), \\
				\mathop{\sup}_j\|\wh\b_j-\b_{j0}\|_2=&O_p((\log p)^{1/2}n^{-1/2}+p^{-1/2}),\\
			\end{split}
		\end{eqnarray*}	
		for each $j=1,\cdots,p$, where $\mathcal{R}_{pn}=n^{-1/2}+p^{-1/2}$.
	\end{theorem}
In fact, the convergence rate  of $\wh\b_j$  consists of two terms, the estimate error $\mathcal{R}_{pn}$ and
	the approximation error $N_0^{-1/2}=n^{-1/2}(n^{-1}\sumi 1/n_i)^{1/2}$, the latter is  from the numerical approximation for $\wt\bSig_{\X}$ and  is ignorable because  $n_i\geq 1$. 
	The convergence rate in Theorem \ref{th1} is similar to those of the linear factor model \citep{bai2013principal} and generalized factor model \citep{liu2021generalized}. 
According to the results of Theorem \ref{th1}, we can further establish the asymptotical normality of the loadings.

	\begin{theorem}\label{th3}
		Under Conditions (A1)-(A6) and when $np^{-2}=o(1),$ for each $j$, we have
		\begin{eqnarray*}
			\sqrt{n}(\wh\b_j-\b_{j0})\d N(0,\LLambda^{-1}_{\bzeta}\PPsi_j \LLambda^{-1}_{\bzeta}),
		\end{eqnarray*}
		where $\PPsi_j=\Ex\Big[\big(\int u_{ij}(t)\PPhi_0(t)dt\big)\bzeta_{i0}
		\bzeta'_{i0}\big(\int\PPhi_0(t)u_{ij}(t)dt
		\big)\Big]$ and $\LLambda_{\bzeta}$ is defined in (\ref{eq:wtSx}).
	\end{theorem}
A similar asymptotical normality for loadings
has also been established in Theorem 2 of  \cite{bai2003inferential}, except that $\PPsi_j$ involves integration over time, which comes from the estimation based on (\ref{eq:B1}).


	\begin{theorem} \label{th2}
		Under Conditions (A1)-(A6), for  $a_1$ defined in Condition (A2) and $\delta$ defined in Condition (A3), we have
		\begin{eqnarray*}
			\hspace{-0.6cm}	\begin{split}
				\|\wh\bzeta_i-\bzeta_{i0}\|_2=&O_p\{\tau_n^{1/2}\mathcal{R}_{pn} + \tau_n^{-r}\}K^{1/2},\\
				\mathop{\sup}_i\|\wh\bzeta_i-\bzeta_{i0}\|_2
				=&
				O_p\Big[\tau^{1/2}_n\big\{
				(\log n)^{1/a_1}n^{-1/2}+n^{1/2\delta}p^{-1/2}\big\} + (\log n)^{1/a_1}\tau^{-r}_n
				\Big]K^{1/2},
			\end{split}
		\end{eqnarray*}	
		for  each $i=1,\cdots,n$.
	\end{theorem}

	Theorem \ref{th2} illustrates the samplewise convergence rate of $\wh\bzeta_i$ consisting of two terms, the estimate error $K^{1/2}\tau_n^{1/2}\mathcal{R}_{pn}$ and approximation error $K^{1/2}\tau_n^{-r}$. 
	When  $\tau_n=O(1)$ and $K=O(1)$, $\PPhi(\cdot)$ is specified by finite parameters, so the model (\ref{eq:model1}) is reduced to the traditional factor model and the estimate error becomes $\mathcal{R}_{pn}$, which is consistent  with the  result of Theorem 1 in \cite{bai2002determining}. 
	In addition, for finite functions with $K=O(1)$ and $n/p\to 0$, the error is controlled by the sample size $n$,
	$\wh\bzeta_i$ reaches the convergence rate of $n^{-r/(2r+1)}$ after taking $v=1/(2r+1)$. The convergence rate comes  from the estimation of finite  eigenfunctions and  is optimal  for  nonparametric function estimation \citep{stone1980optimal}.
	Theorem \ref{th2} also shows that the uniform convergence rate of $\wh\bzeta_i$ is slower than the  samplewise  convergence rate.  This conclusion is consistent with those for the linear factor model \citep{bai2013principal} and generalized factor model \citep{liu2021generalized}.  In particular, when $\tau_n=O(1)$ and $K=O(1)$ for the parametric factor model,  the uniform estimate error $\left\{
	(\log n)^{1/a_1}n^{-1/2}+n^{1/2\delta}p^{-1/2}
	\right\}\tau^{1/2}_nK^{1/2}$  reduces to $(\log n)^{1/a_1}n^{-1/2}+n^{1/2\delta}p^{-1/2}$,  which is similar to  the uniform rate $n^{-1/2}+n^{1/2\delta}p^{-1/2}$, as shown in Theorem 3.1 in \cite{bai2013statistical} for the  traditional factor model. 

	\begin{rem} To  guarantee the  samplewise consistency of factors or variablewise consistency of loadings, we require that $p$  diverges at any rate, including 
	an exponential rate with respect to
	$n$. Our simulation studies  show that the performance of 
	the proposed method improves as $p$ and $n$ increase. That is, the high dimensional $p$ is a blessing of dimensionality instead of a curse. This is a direct result of the
	factor model, where $p$ actually plays the role of the number of observations for estimating $\bzeta_i$.
	In our opinion,
	more variables provide more information to estimate the models when the factor number or latent freedom is fixed.  The issue 
	where the high dimensional $p$ is a blessing of dimensionality instead of 
	curse has also been well addressed  for the linear factor model in  \cite{li2018embracing}.  
	\end{rem}

	\begin{theorem}\label{thphi}
		Under Conditions (A1)-(A6)  for each $j=1,\cdots,q$, we have
		\begin{eqnarray*}
			\sum_{k=1}^{K}\int\left(\wh\phi_{jk}(t)-\phi_{jk,0}(t)\right)^2dt=O_p\{\tau_n\mathcal{R}^2_{pn} + \tau_n^{-2r}\}K
		\end{eqnarray*}
	\end{theorem}
	
	Theorem \ref{thphi} establishes the convergence rate  for eigenfunctions $\wh\PPhi(t)$, which consists of two terms, the estimate error $K^{1/2}\tau^{1/2}_n\mathcal{R}_{pn}$ and approximation error $K^{1/2}\tau^{-r}_n$, which is
	similar to Therorem \ref{th2}.

\section{Numerical Studies}
\label{sec:simu1}

In this section, we conduct simulation studies to assess the finite-sampling performance of the proposed method (FaFPCA) by comparing it with state-of-the-art methods, 
including uFPCA-HD  \citep{gao2019high},  FM-HD \citep{tavakoli2021high},  sparse functional principal analysis (sFPCA-HD, \citealt{hu2022sparse}),  dynamic principal component analysis (DPCA, \citealt{hu2021dynamic}) for  high-dimensional functional data. 
The performance of the estimator is evaluated via  the computational time and the accuracy of the estimator, including the root mean square error RMSE$_l=p^{-1/2}\|\wh\B-\B_0\|_F$ for the loadings,  RMSE$_f=n^{-1/2}\|\wh\bzeta-\bzeta_0\|_F$ for the factors,
	RMSE$_e=\|\wh\PPhi(t)-\PPhi(t)\|_F=\left(\sum_{j,k}\int(\wh\phi_{jk}(t)-\phi_{jk0}(t))^2\ dt
	\right)^{1/2}$ for the eigenfunctions, 	and normalized prediction error (PE)$=\sum_{i,j}n^{-1}_i\sum_{l=1}^{n_i}(\wh X_{ij}(t_{il})-X_{ij}(t_{il}))^2/\sum_{i,j}n^{-1}_i\sum_{l=1}^{n_i} X^2_{ij}(t_{il})$ for the observations $X_{ij}(t_{il})$ in the testing set, which is independent of the training data but has the same distribution and size as the training data. 

\subsection{Simulation setting}

For  each trajectory $\X_i(t)$,  $20$  observation time points are randomly sampled from $U(0,10)$ and 200 replicates are applied in each scenario unless otherwise stated.

{\bf Scenario 1:} To be fair, we generate data from the following model, which does not follow the assumptions of the methods we consider, $$\X_i(t)=\sum^{K}_{k=1}\bzeta_{ik}\phi_k(t)+\bm{\epsilon_i}(t),\ i=1,\cdots, n,$$
where  $\bzeta_{ik}\in\RX^p\sim N(\0,\bSig_{\bzeta,k})$,  $\bm{\epsilon}_i(t)\sim N(\0,\bSig)$ for each $t$,
$(\bSig_{\bzeta,k})_{ij}=0.5^{|i-j|}$ if $i\neq j$ and  $(\bSig_{\bzeta,k})_{ii}=3-(k-1)/(K-1)$  for each $i=1,\cdots, p$ and $k=1,\cdots,K$.
We take
$\bSig=(\sigma_{ij})_{p\times p}$  as ${\bf Case\  \Rmnum{1}}:\bSig=\I$ and ${\bf Case\  \Rmnum{2}}: \sigma_{ii}=1$ and $\sigma_{ij}=0.3$ if $i\neq j$,
and let $\phi_k(t)=\sin[(2k-1)\pi t/10]$,  $n=p=100$ and  $K=2, 10$. 

{\bf Scenario 2:} To investigate the performance of the estimation for loadings, factors and eigenfunctions, we generate functional data  by $\X_i(t) = \B \h_i(t) + \u_i(t), \ i=1,\cdots, n,$  which satisfie the assumption of the proposed method, where random error $\u_i(t)\sim N(\0,\I)$.
To construct $\B$, we first generate $n$ samples of the $p$-dimensional $\k_i$ from $N(0, (\sigma_{ij})_{p\times p})$  with $\sigma_{ij}=0.5^{|i-j|}$. Here, $\K=(\k_1,\cdots,\K)'.$ We apply eigendecomposition to  the matrix $\K\K'$ 
to obtain $\B_{q}$ and $\K_q$, where $\B_{q}$ is a diagonal matrix whose diagonal entries are the first $q$ largest eigenvalues of $\K\K'$, and $\K_{q}\in\RX^{n\times q}$ consists of the corresponding eigenvectors.	
$\B_n=\K'\K_{q}$, and QR decomposition of  $\B_n=\Q_n\R_n$ is performed. Then, $\B=p^{1/2}\Q_n$ so that the identification condition on $\B$ can be satisfied.
Similarly, to construct $\bzeta_i$ for $\h_i(t)$, we independently generate $\xi^{*}_{ik}\sim N(0, Kqk^{-1})$, denote $\bzeta^{*}_i=(\xi^{*}_{i1},\cdots,\xi^{*}_{i,Kq})'$ and $\bzeta^{*}=(\bzeta^{*}_1,\cdots,\bzeta^{*}_n)$, and then  perform eigen-decomposition of the matrix  $\bzeta^{*\prime}\bzeta^{*}$ for  matrices $\LLambda_{Kq}$ and  $\M_{Kq}$, which consist of the first $Kq$ largest eigenvalues and the corresponding eigenvectors, respectively. We  take  $\bzeta\hat=(\bzeta_1,\cdots,\bzeta_n)'=\LLambda^{1/2}_{Kq}\M'_{Kq}.$  It is easy to check that $\bzeta$ satisfies Identification Condition (I2), as described in Section \ref{model}.
Given $\bzeta_{i},$ the factor process $\h_i(t)=(h_{i1}(t),\cdots,h_{iq}(t))^\prime$ is generated by $h_{ij}(t)=\sum_{k=1}^K   \xi_{ijk}\phi_{jk}(t)$, where   $\phi_{jk}(t)=\sqrt{2}\sin[\{2(j\neq 2)+(j=2)\}k\pi t/10]$ if $k$ is odd and
$\phi_{jk}(t)=\sqrt{2}\cos[\{2(j\neq 2)k+(j=2)(2k+1)\}\pi t/10]$ if $k$ is even. The setting is used to ensure the  orthogonality of eigenfunctions.


\subsection{Simulation results}\label{sec:simu2}
	
{\bf Scenario 1:}  We first compare FaFPCA with the existing methods in terms of prediction accuracy PE.
Figure \ref{model2} shows 
the PE
of the proposed FaFPCA, FM-HD, DPCA, sFPCA-HD  and  uPFCA-HD 
methods with the corresponding optimal tuning
parameters under the cases listed in Scenario 1. From Figure \ref{model2}, we can see  that
(1) When $\u_i(t)$ becomes correlated, that is, when $\bSig$ changes from $\I$ to $\bSig\neq \I$ in  Figures \ref{model2}(a) and \ref{model2}(b), respectively,  FaFPCA, FM-HD, and DPCA perform better, while the PE performance of sFPCA-HD  and  uPFCA-HD 
worsens. This indicates that  FaFPCA, FM-HD and DPCA are efficient in making use of the correlations among functional variables
while sFPCA-HD  and  uPFCA-HD  are not. 
(2) On the other hand, when $K$ changes from 10 to 2 in Figures \ref{model2}(c) and \ref{model2}(d), which means the dependence over time gets stronger,  sFPCA-HD  and  uPFCA-HD perform better,  while  FM-HD and DPCA  
perform
worse. This is not surprising  because sFPCA-HD  and  uPFCA-HD focus on  dependence over time
while FM-HD and DPCA do not.
(3) The proposed  FaFPCA  approach performs the best and is stable in all  the cases in terms of PE; particularly, either  the dependence over time or correlations among variables increase, the accuracy of FaFPCA is improved.
Compared to the other methods, the 
obtained by
FaFPCA is due to the usage of  the two kind of  correlations.

	\begin{figure}[!htb]
		\begin{center}
		\subfigure[$K=10,$ $\bSig=\I$(red),  $\bSig\neq\I$(blue)]{
			\includegraphics[width=0.4\textwidth,height=0.4\textwidth]{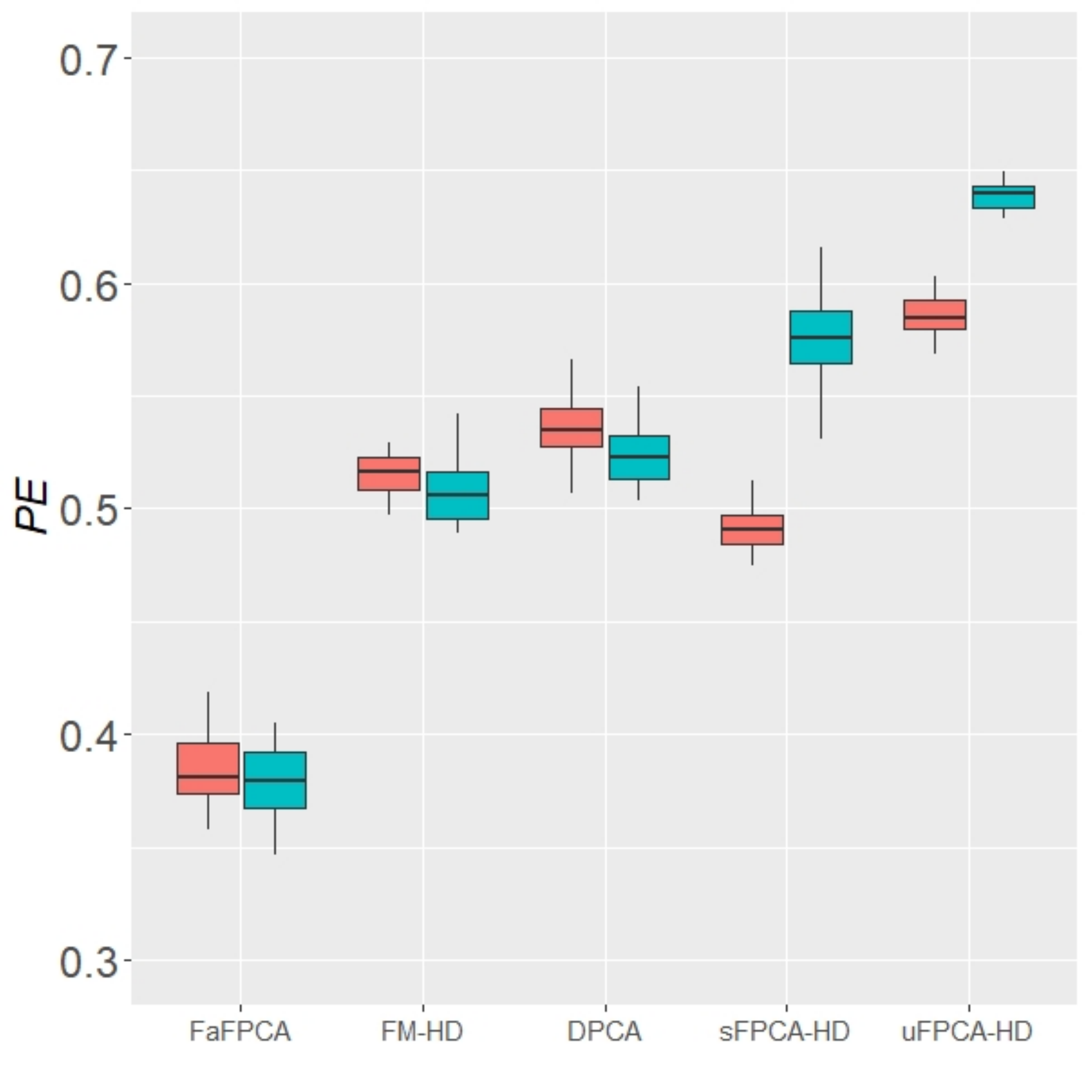}}
		\subfigure[$K=2,$ $\bSig=\I$(red),  $\bSig\neq\I$(blue)]{
			\includegraphics[width=0.4\textwidth,height=0.4\textwidth]{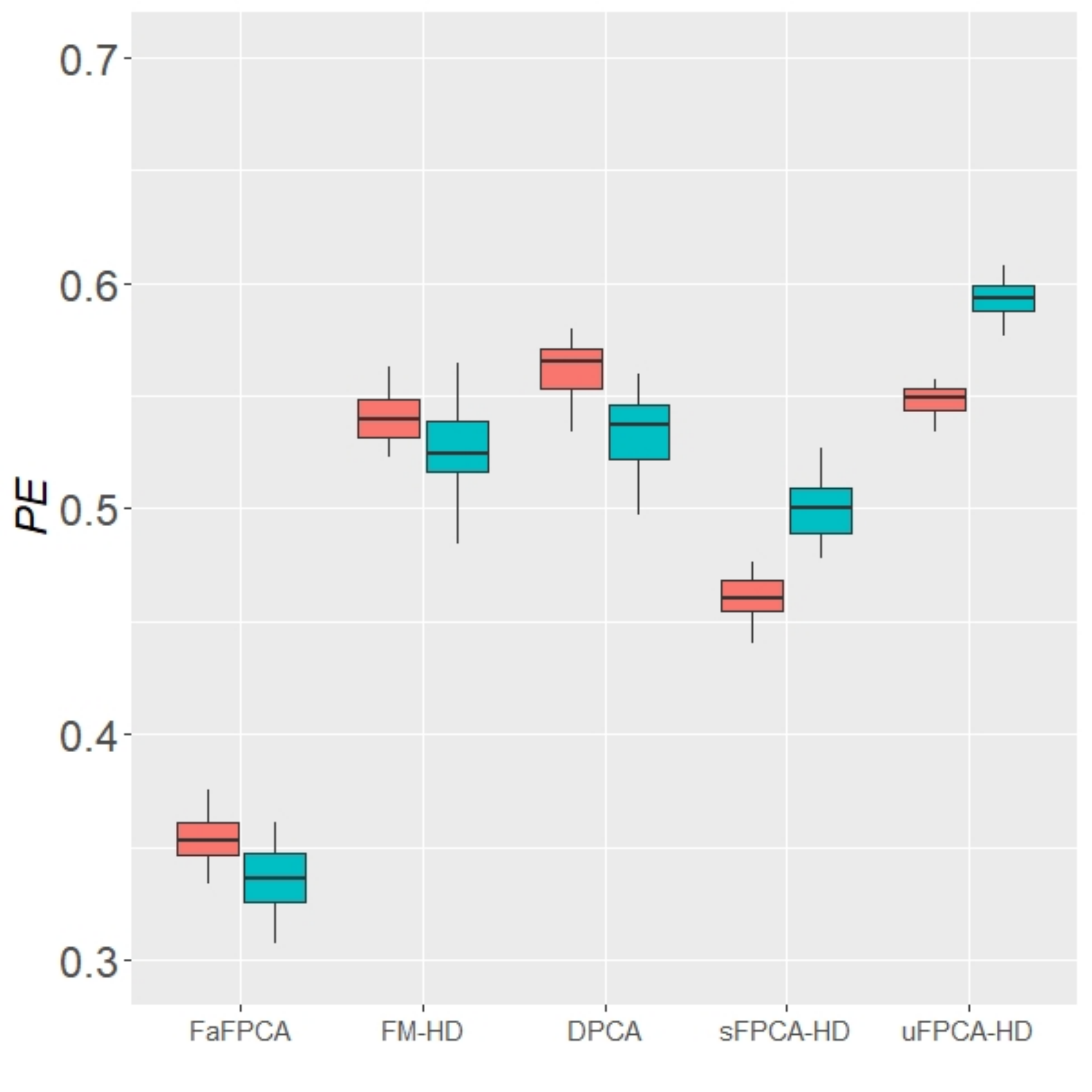}}
		\\
		\subfigure[$\bSig=\I,$ $K=10$(red) and $K=2$(blue)]{
			\includegraphics[width=0.4\textwidth,height=0.4\textwidth]{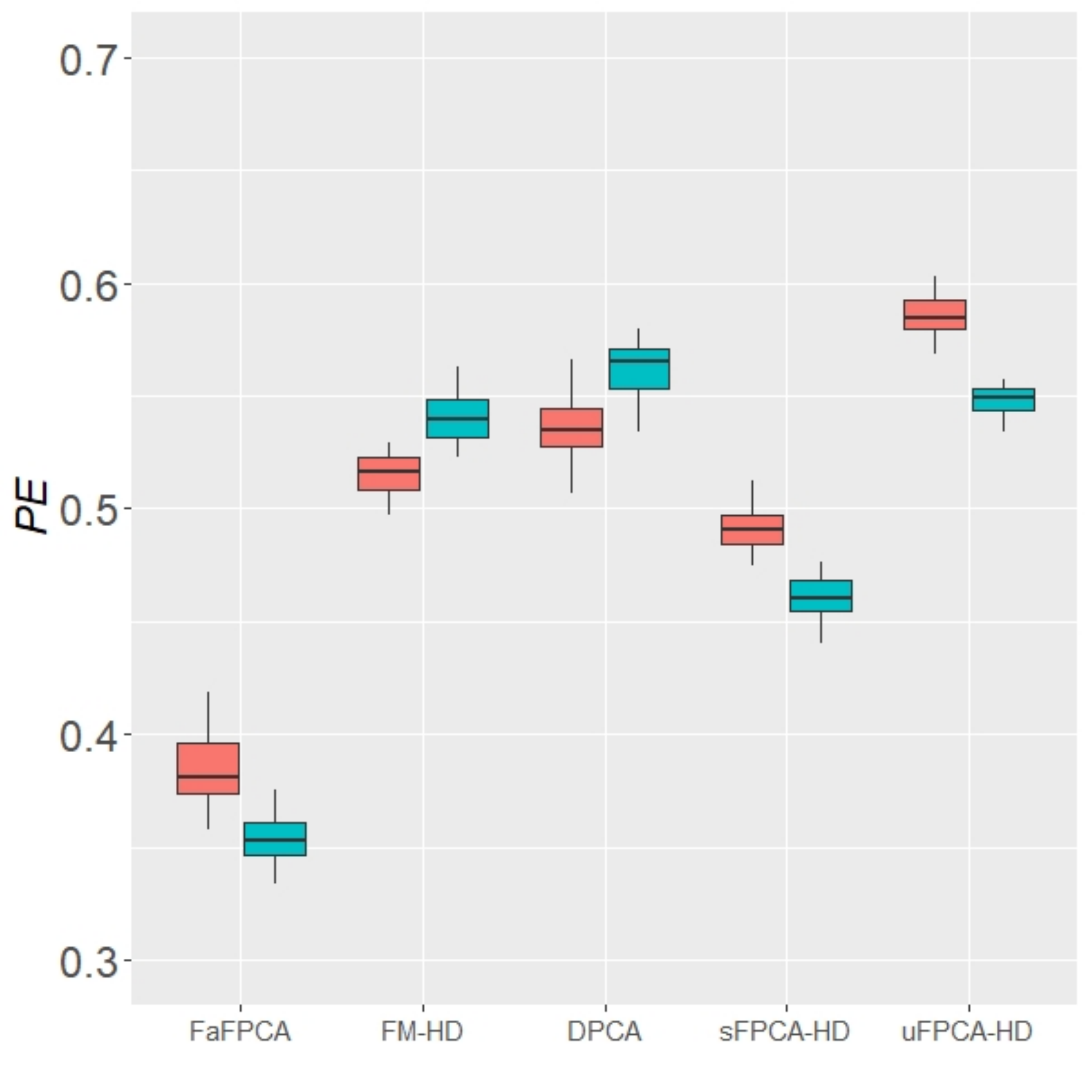}}
		\subfigure[$\bSig\neq\I,$ $K=10$(red) and $K=2$(blue)]{
			\includegraphics[width=0.4\textwidth,height=0.4\textwidth]{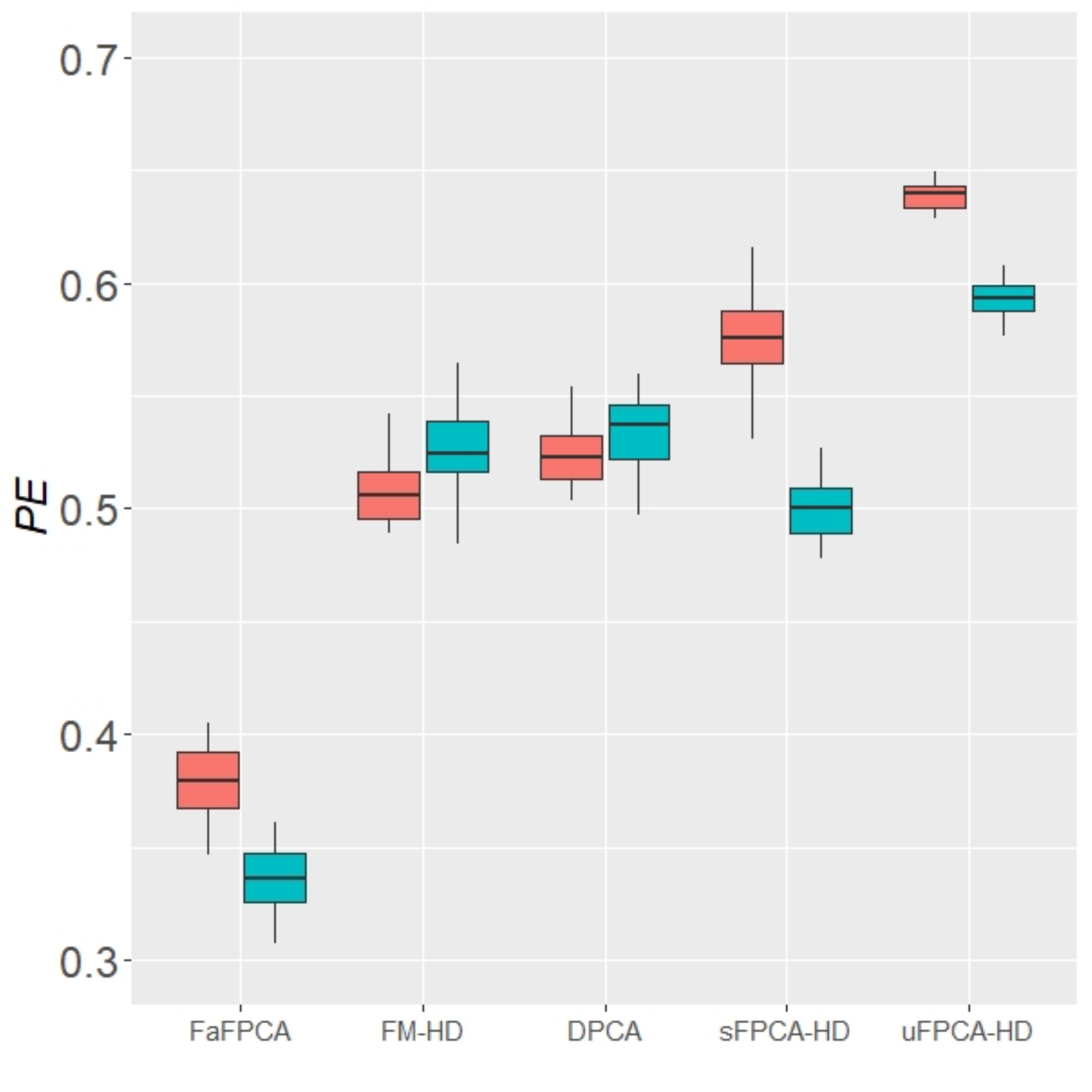}}
		\end{center}
		\caption{The PEs of the FaFPCA, FM-HD, DPCA, sFPCA-HD and uPFCA-HD methods under the cases listed in Scenario 1. }
		\label{model2}
	\end{figure}

We carried out the computation on  a single 4-core personal laptop with 16 GB of RAM.  Table \ref{table:time} shows  the average running time of 200 repetitions for Case  \Rmnum{1} with $K=2$.  
Table \ref{table:time} shows that the proposed FaFPCA approach is much faster than  the other methods except FM-HD, which  only considers model (\ref{eq:bh}) and does not estimate eigenfunctions and factor scores. uFPCA-HD requires 
an extremely large amount of time because uFPCA is applied
to each of the $p$ curves separately. 

\begin{table}[!htb]
	\small
	\centering
	\caption{The computational time in seconds of the five  methods.}\label{table:time}
	\renewcommand\arraystretch{1.7}
	\begin{center}
	\begin{tabular}{r|rrrrr}
		\hline
		\rule{0pt}{11pt}			
		&FaFPCA&FM-HD&DPCA&sFPCA-HD&uFPCA-HD\\
		\hline
		Mean&0.0300&0.0190 &0.4570  &0.2430   &77.3500\\
		SD  &0.0097&0.0095 &0.0178  &0.0283   &3.0647\\
		\hline
	\end{tabular}
	\end{center}
\end{table}

{\bf Scenario 2:}  Figure \ref{rmse}  displays  RMSE$_l$ for loadings,  RMSE$_f$ for factors and RMSE$_e$  for eigenfunctions using the proposed FaFPCA method for Scenario 2 with  $(q,K)=(5,2)$ when $(n,p)$ varies. 
The results in Figure \ref{rmse} indicate that the precision of $\B, \bzeta$ and $\PPhi(t)$ increases as $n$
or $p$ increases, which is consistent with the theoretical results. In particular,  the precision of  $\bzeta$ is  more sensitive to $p$ than to $n$, while that of $\B$  is  more sensitive to $n$.  This is 
because the estimator for $\bzeta_i$ is based on $p$ functions of individual $i$; however, the estimator for $\B$ is mainly based on  $n$ individuals.

\begin{figure}[!htb]
	\begin{center}	
	\begin{minipage}[c]{0.32\linewidth}
		\centering
		\includegraphics[height=5.2cm,width=5.2cm]{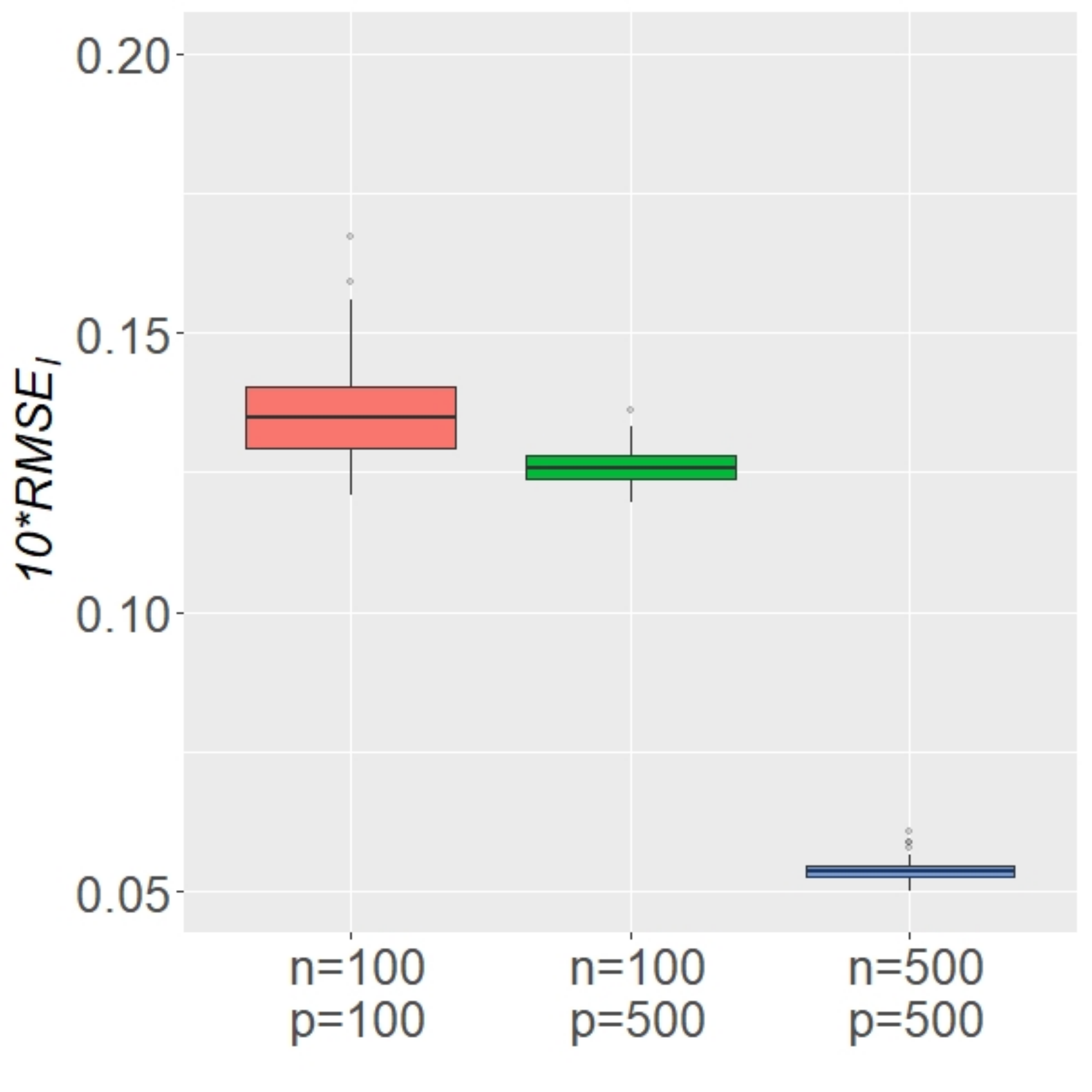}
	\end{minipage}%
	\begin{minipage}[c]{0.32\linewidth}
		\centering
		\includegraphics[height=5.2cm,width=5.2cm]{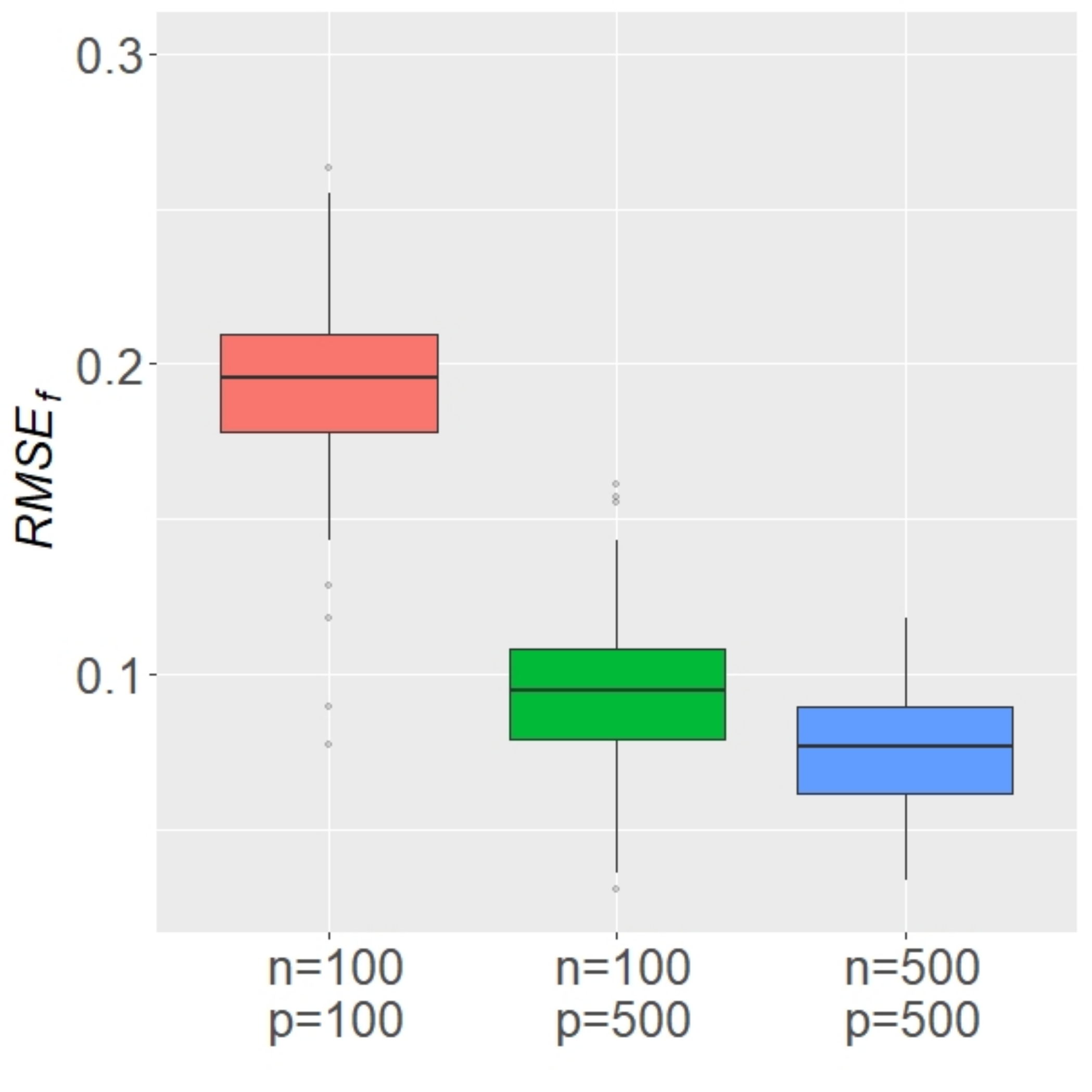}
	\end{minipage}%
	\begin{minipage}[c]{0.32\linewidth}
		\centering
		\includegraphics[height=5.2cm,width=5.2cm]{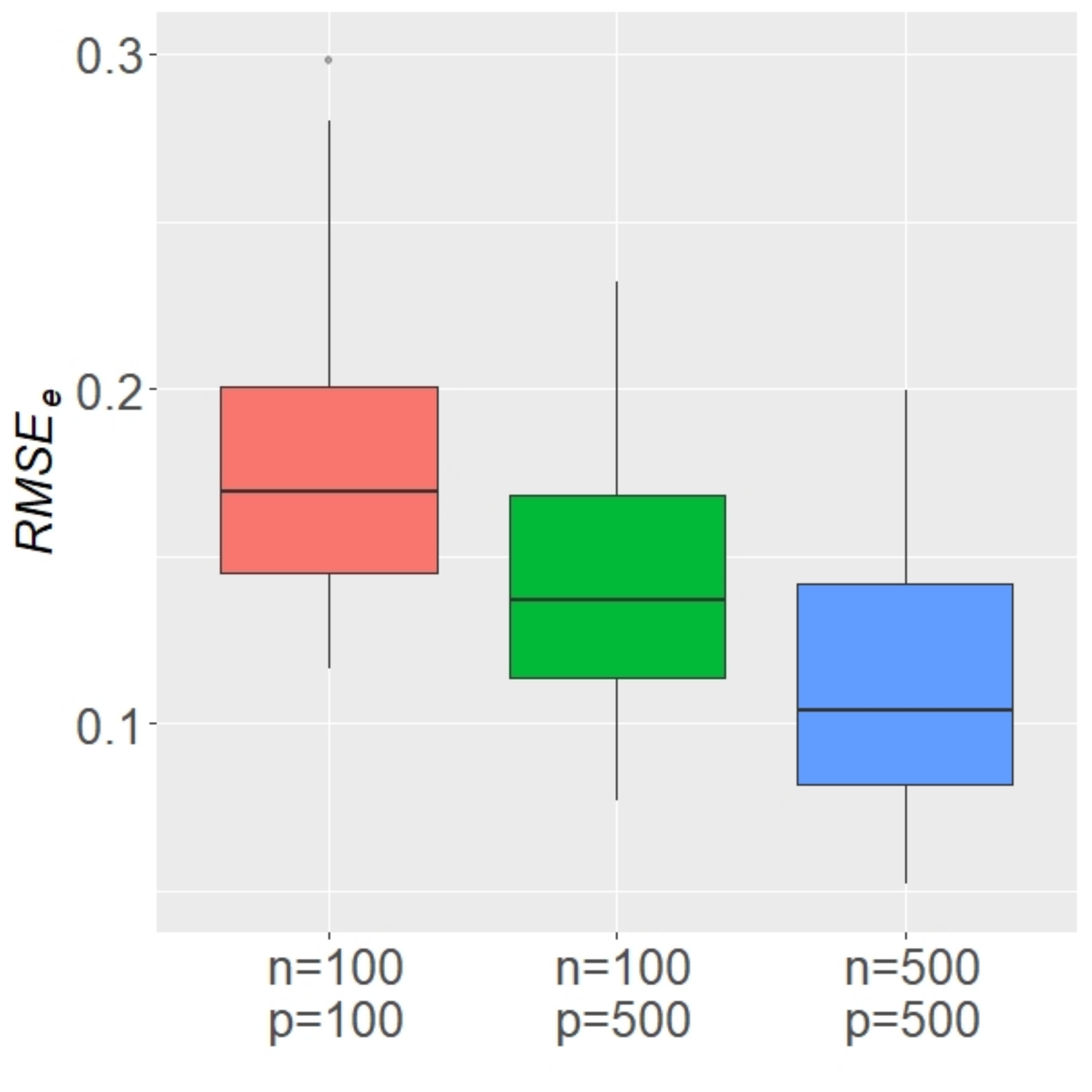}
	\end{minipage}%
	\end{center}

	\caption{The boxplots of $10\times$RMSE$_l$,RMSE$_f$ and RMSE$_e$ (from left to right) of the FaFPCA for Scenario 2 with $n, p=100,500$. The RMSE$_l$, RMSE$_f$ and RMSE$_e$ are the root of mean
square errors for loadings, factors and eigenfunctions, respectively.  }\label{rmse}
\end{figure}


To show the effect of the number of observations, we take  $n_i$ from 5 to 50 given  $n=p=100$ and $(q,K)=(5,2)$. 
The results in Figure  \ref{rmse_t} show that the estimators are quite stable with respect to $n_i$ except for small $n_i$  values.  When $n_i$ is small,  the matrix $\sum_{l=1}^{n_i}\M(t_{il})\M'(t_{il})$  in $\w_{ik}$ might be  irreversible.
To avoid this problem, we  add a  term $\lambda\I_{\tau_n}$, 
where
$\lambda$ is small, to $\sum_{l=1}^{n_i}\M(t_{il})\M'(t_{il})$  to ensure invertibility. This may cause extra bias for the estimations of factors and eigenfunctions when $n_i$ is small.
From Figure  \ref{rmse_t}, we also observe an interesting phenomenon as $n_i$ increases, RMSE$_{l}$ for loadings increases, while   RMSE$_{f}$ and  RMSE$_{e}$ for factors and eigenfunctions, respectively, decrease.
This can be attributed the following two aspects.  On the one hand,
the loadings are estimated mainly based on $n$ samples, rather than $n_i$ or $p$; hence, the increasing $n_i$ values may introduce extra noise due to the integration of all observations using (\ref{eq:B11}).
On the other hand,  the factors $\bzeta_i$ are estimated mainly based on the observations from individual $i$; hence, the estimation of the factors becomes better as $n_i$ increases, although  becomes stable when $n_i$ is large enough, for example, when $n_i>10$ in the cases. The estimation 
of eigenfunctions has a similar performance as that of the factors since they are simultaneously estimated.

\begin{figure}[!htb]
	\begin{center}	
	\begin{minipage}[c]{0.32\linewidth}
		\centering
		\includegraphics[height=5.2cm,width=5.2cm]{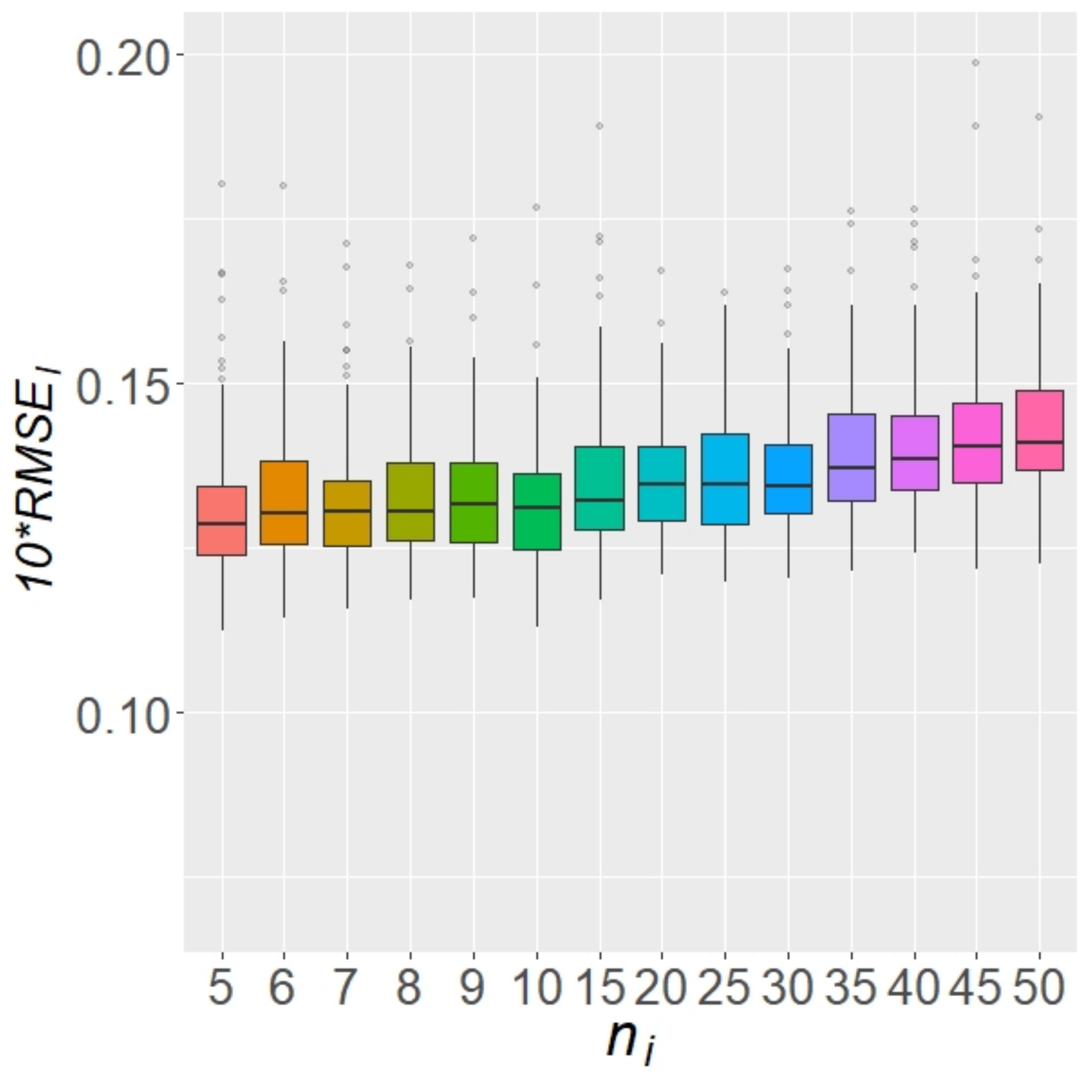}
	\end{minipage}%
	\begin{minipage}[c]{0.32\linewidth}
		\centering
		\includegraphics[height=5.2cm,width=5.2cm]{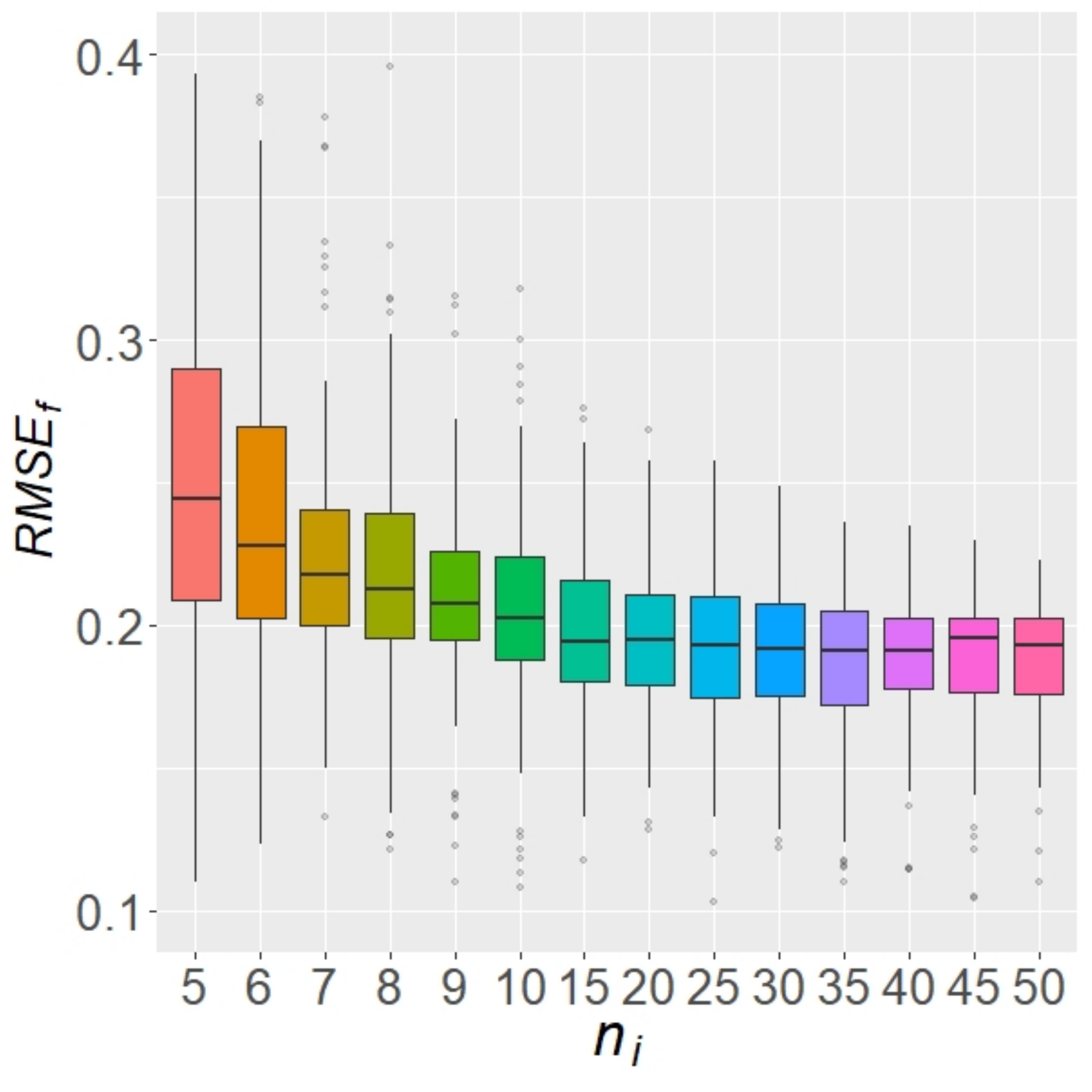}
	\end{minipage}%
	\begin{minipage}[c]{0.32\linewidth}
		\centering
		\includegraphics[height=5.2cm,width=5.2cm]{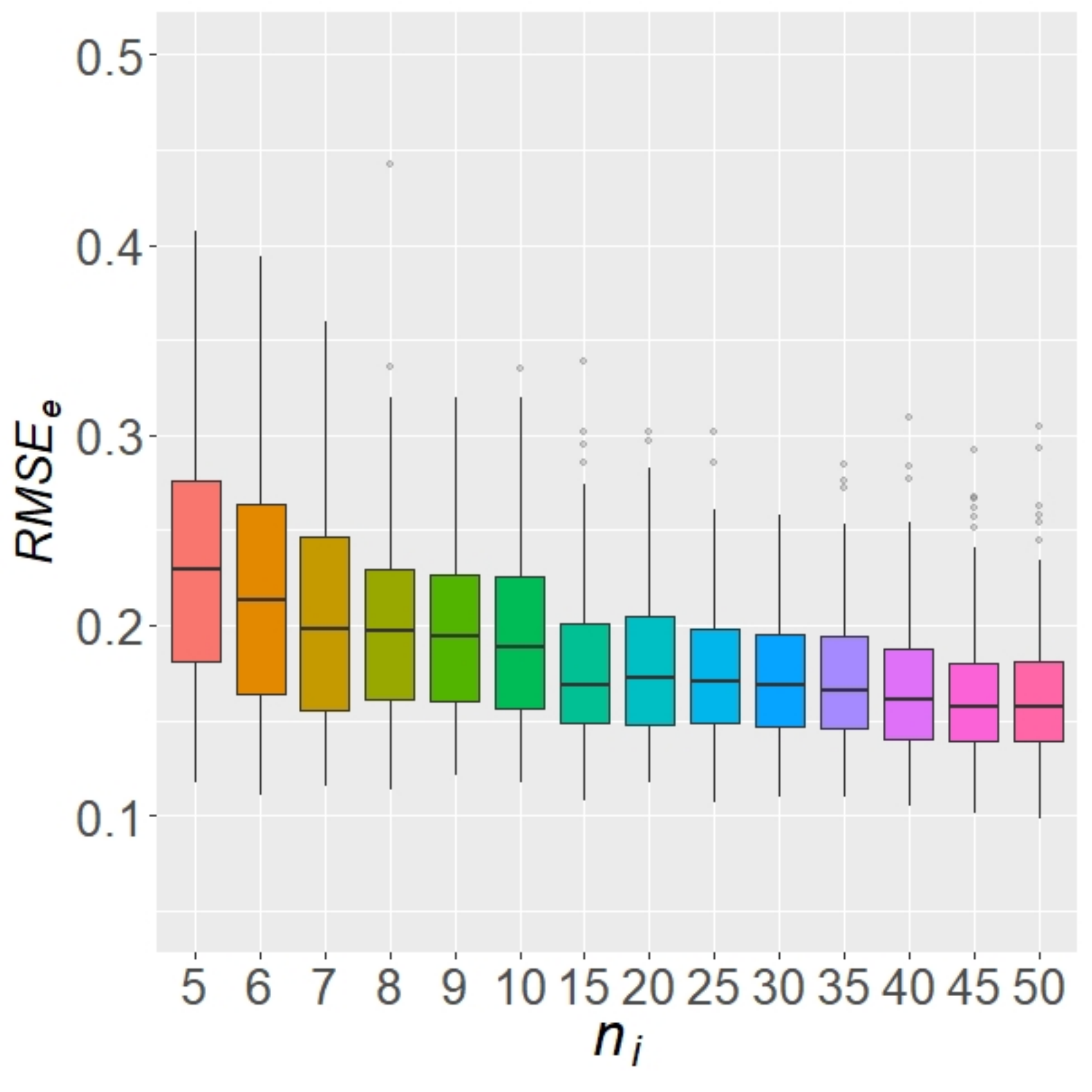}
	\end{minipage}%
	\end{center}
	
	\caption{The boxplots of $10\times$RMSE$_l$, RMSE$_f$ and RMSE$_e$ (from left to right) of the FaFPCA when $n_i$ changes from 5 to 50. The RMSE$_l$, RMSE$_f$ and RMSE$_e$ are the root of mean
square errors for loadings, factors and eigenfunctions, respectively.  }\label{rmse_t}
\end{figure}


\section{Real Data Analysis}

Alzheimer’s disease (AD) is irreversible and the most common form of dementia, and it
can result in the loss of thinking, memory and language skills. It is important to 
determine the progress of AD 
by monitoring alterations in the brain and to
detect the disease at an early stage.  Particularly, it is well known that the atrophy of the region of interest (ROI) is associated with AD detection and some studies consider local brain volumes for characterizing regional atrophy \citep{zhao2019heritability,zhong2021cluster,li2022asynchronous}.
In the paper,
we use the brain volume density values within different ROIs  to detect the progression of AD. 
The dataset includes a subset of 
participants with 766 subjects enrolled in the first phase of the ADNI study \citep{mueller2005alzheimer} with AD, mild cognitive impairment
(MCI, an early stage of AD) and cognitively normal (CN). 
A total of 172, 378 and 216 participants were diagnosed with AD, MCI and CN, respectively.
Each participant’s
record consists of the brain volume density values of 90 ROIs for each of  the 501 equispaced sampling volumes in the interval of $[-255,\ 255]$, so that we can regard the brain volume density as a curve.  
To be specific, we denote $\X_i(t)$ by the brain volume density curves of ROIs of the $i$-th individual, which are functions of the log of the Jacobian volume (denoted by $t$). We scale the log Jacobian volume into $[0,\ 1]$ before analysis and the summary of the ROIs is
shown in Table \ref{table:cluster}.
Consequently,  90 ROIs are  high-dimensional functional variables that can be  analyzed by the proposed FaFPCA approach, as well as the methods mentioned in Section \ref{sec:simu1} for comparison. 

We  apply FaFPCA, FM-HD, DPCA, sFPCA-HD and uFPCA-HD to the  data without the label information of AD, MCI and CN.  First, based on the criteria in Section \ref{section:2.3}, we choose $q=5$ factors and $K=3$ eigenfunctions for FaFPCA. Similarly, on the grounds of the proportion of variability, 
we choose  $q=5$, $K=15$, $K=33$ and $K=2$ for the FM-HD, DPCA, sFPCA-HD and uFPCA-HD methods, respectively.
We randomly set 60\% of the dataset as the training  set and the other data as the test set with 200 repeats. 
Figure \ref{npex_realdata}(a) presents the PEs of FaFPCA, FM-HD,  DPCA, sFPCA-HD and uFPCA-HD with the corresponding selected hyperparameters and shows that FaFPCA yeilds the  estimator with the highest prediction accuracy.
In addition, Figure \ref{npex_realdata}(a)  also shows  that 
the sFPCA-HD and uFPCA-HD methods are the worst in terms of PE, which indicates that the correlation among functional variables might  be strong in the ADNI data.

	\begin{figure}[!htb]
		\begin{center}
		\subfigure{(a)
			\includegraphics[width=0.35\textwidth,height=0.35\textwidth]{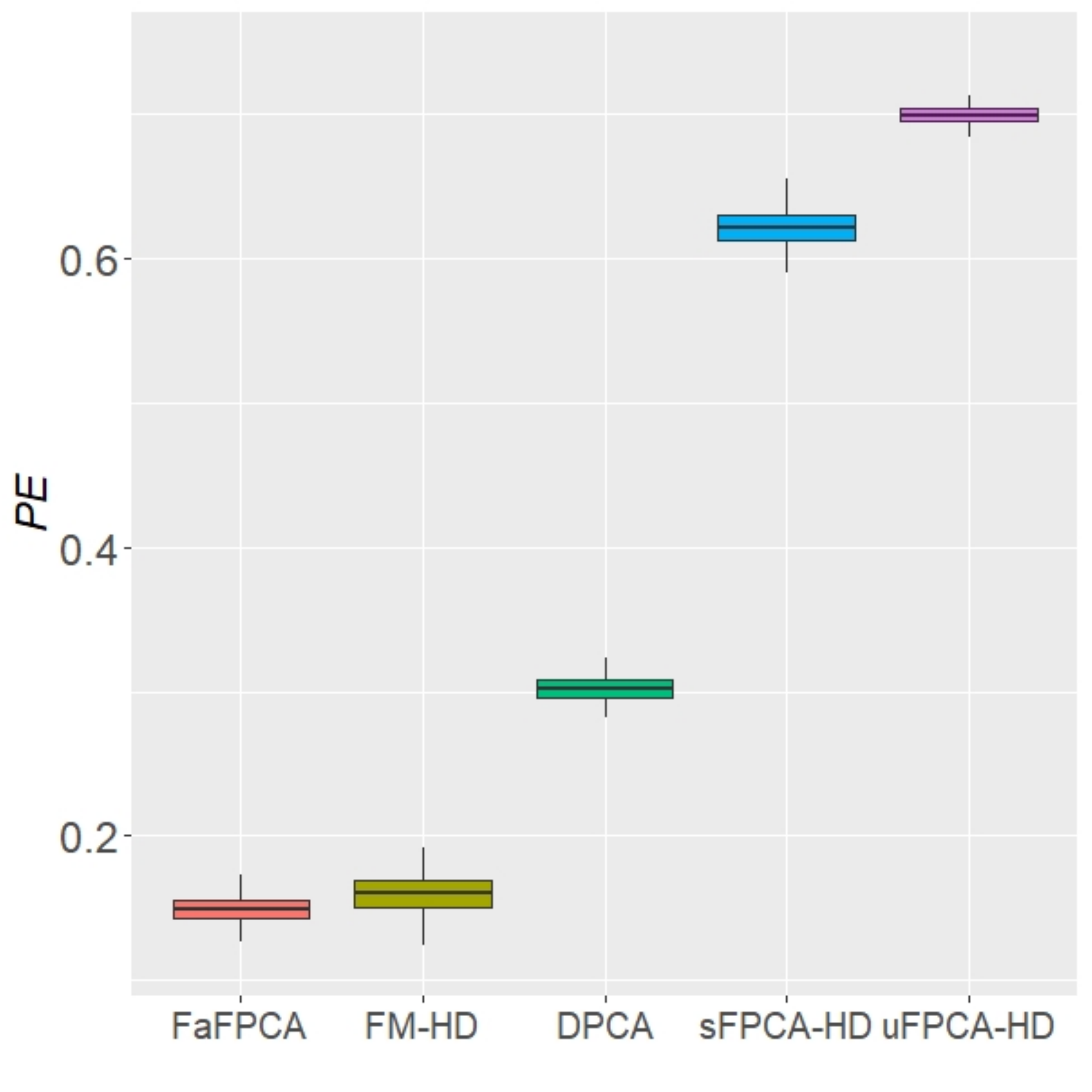}}
				\subfigure{(b)
			\includegraphics[width=0.45\textwidth,height=0.35\textwidth]{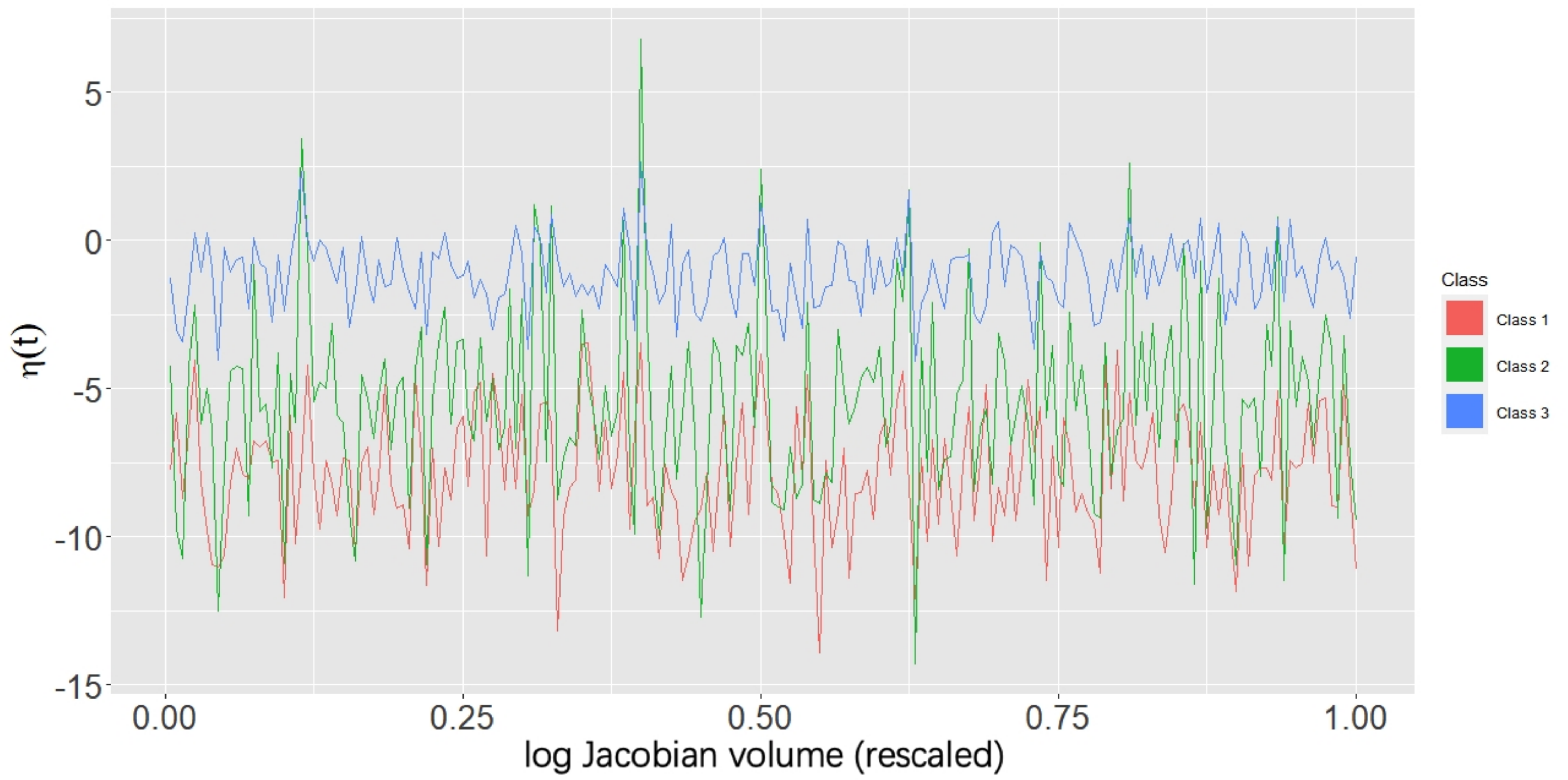}}
		\end{center}
		\caption{(a) The PEs of  FaFPCA, FM-HD, DPCA,  sFPCA-HD and uFPCA-HD for the ADNI dataset.
		(b)  The regression coefficient functions of choosed ROIs in each class.}
		\label{npex_realdata}
	\end{figure}

Now, we investigate  details about the factors in Figure \ref{loading_realdata}. Based on  the top ten largest absolute values of the loadings,  factor 1 extracts the  feature mainly from
variables ROI$_{12}$, ROI$_{15}$, ROI$_{16}$, ROI$_{51}$, ROI$_{91}$, ROI$_{92}$, ROI$_{631}$, ROI$_{632}$, ROI$_{1016}$ and ROI$_{2016}$, which represent the information on the brain regions starting from the forebrain, arriving at the base of the brain, and then 
returning to the afterbrain. These ROIs  
exist in the bottom lateral part of the brain and  
mostly in the cerebellum.
Factor 2 withdraws the information of variables ROI$_{5}$, ROI$_{16}$, ROI$_{17}$, ROI$_{18}$, ROI$_{28}$, ROI$_{50}$, ROI$_{53}$, ROI$_{54}$, ROI$_{49}$ and ROI$_{60}$,
which express information on the basal nucleus and are located deeply in the white matter of the brain. Factors 3-5 extract the  information 
from regions similar to those of factor 2.

\begin{figure}[!htb]
\begin{center}
        \subfigure{
			\includegraphics[width=0.3\textwidth,height=0.3\textwidth]{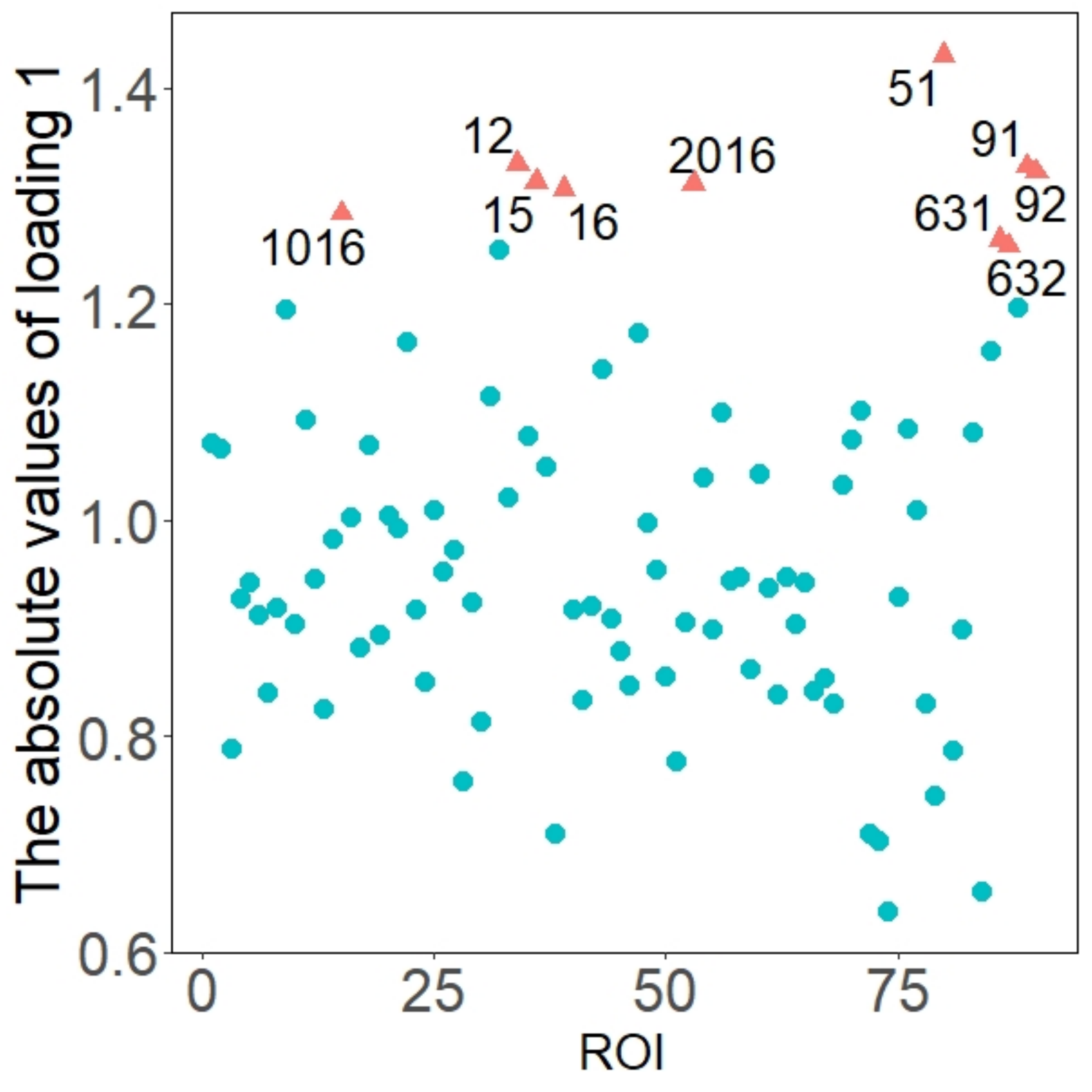}}
		\subfigure{
			\includegraphics[width=0.3\textwidth,height=0.3\textwidth]{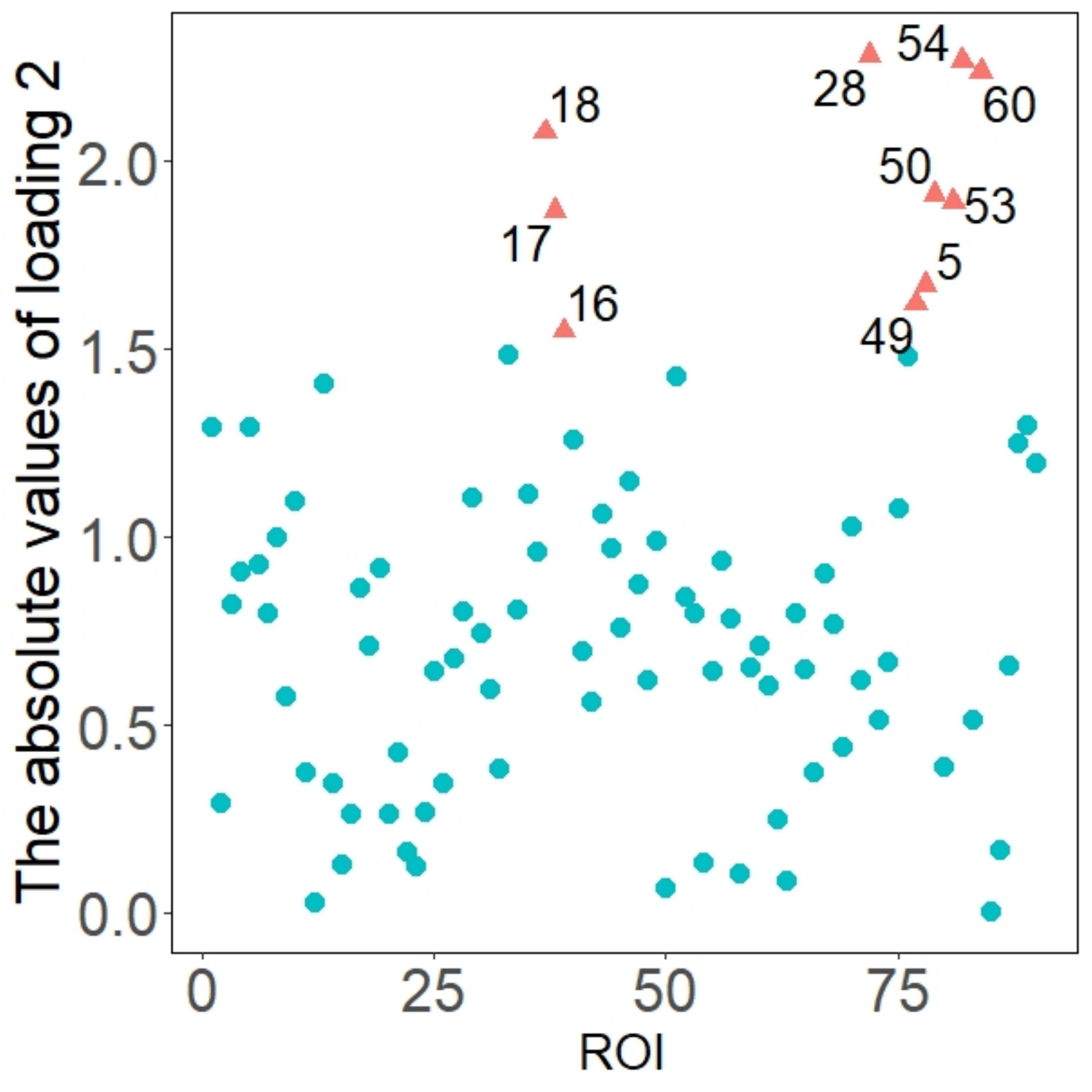}}
		\subfigure{
			\includegraphics[width=0.3\textwidth,height=0.3\textwidth]{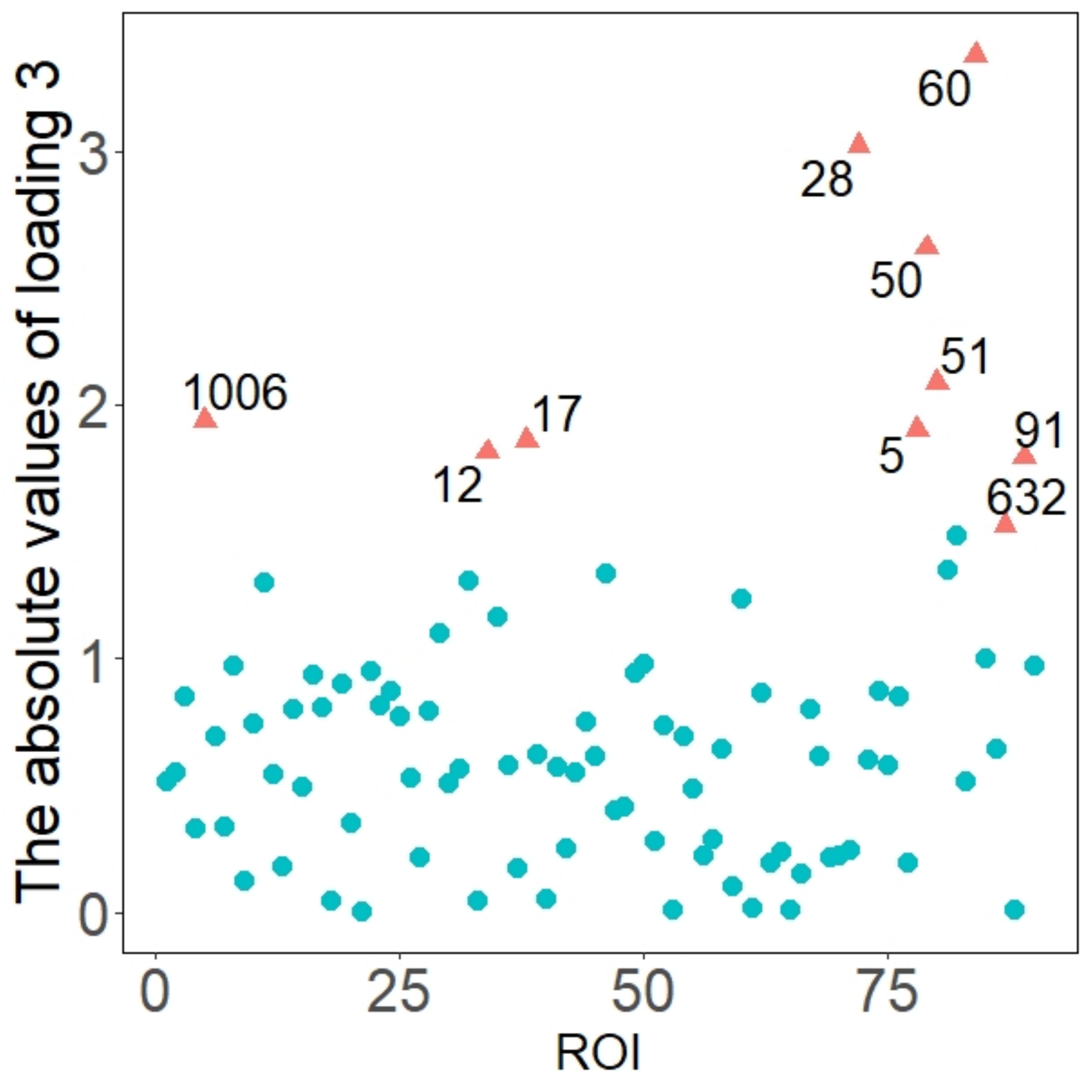}}
		\subfigure{
			\includegraphics[width=0.3\textwidth,height=0.3\textwidth]{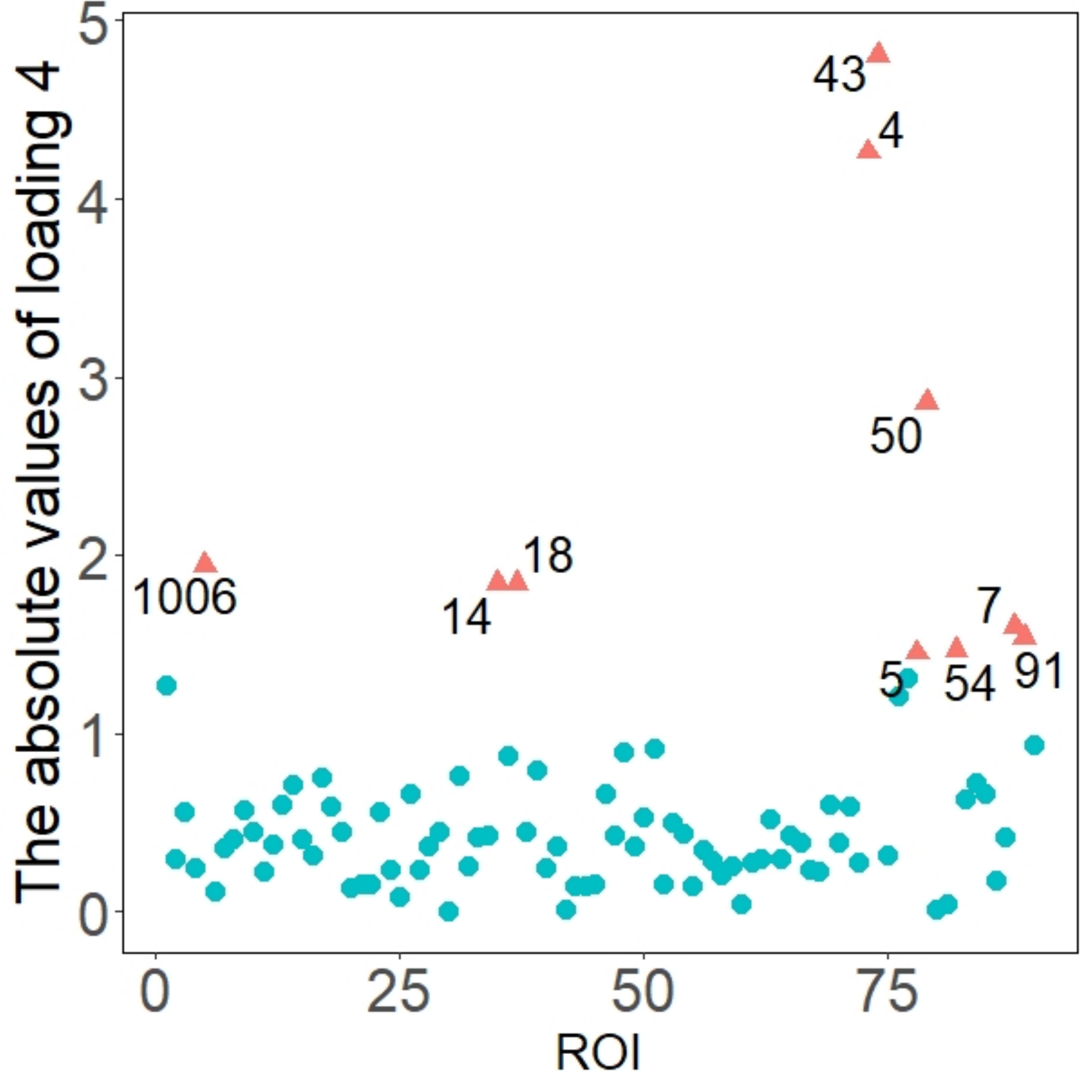}}
		\subfigure{
			\includegraphics[width=0.3\textwidth,height=0.3\textwidth]{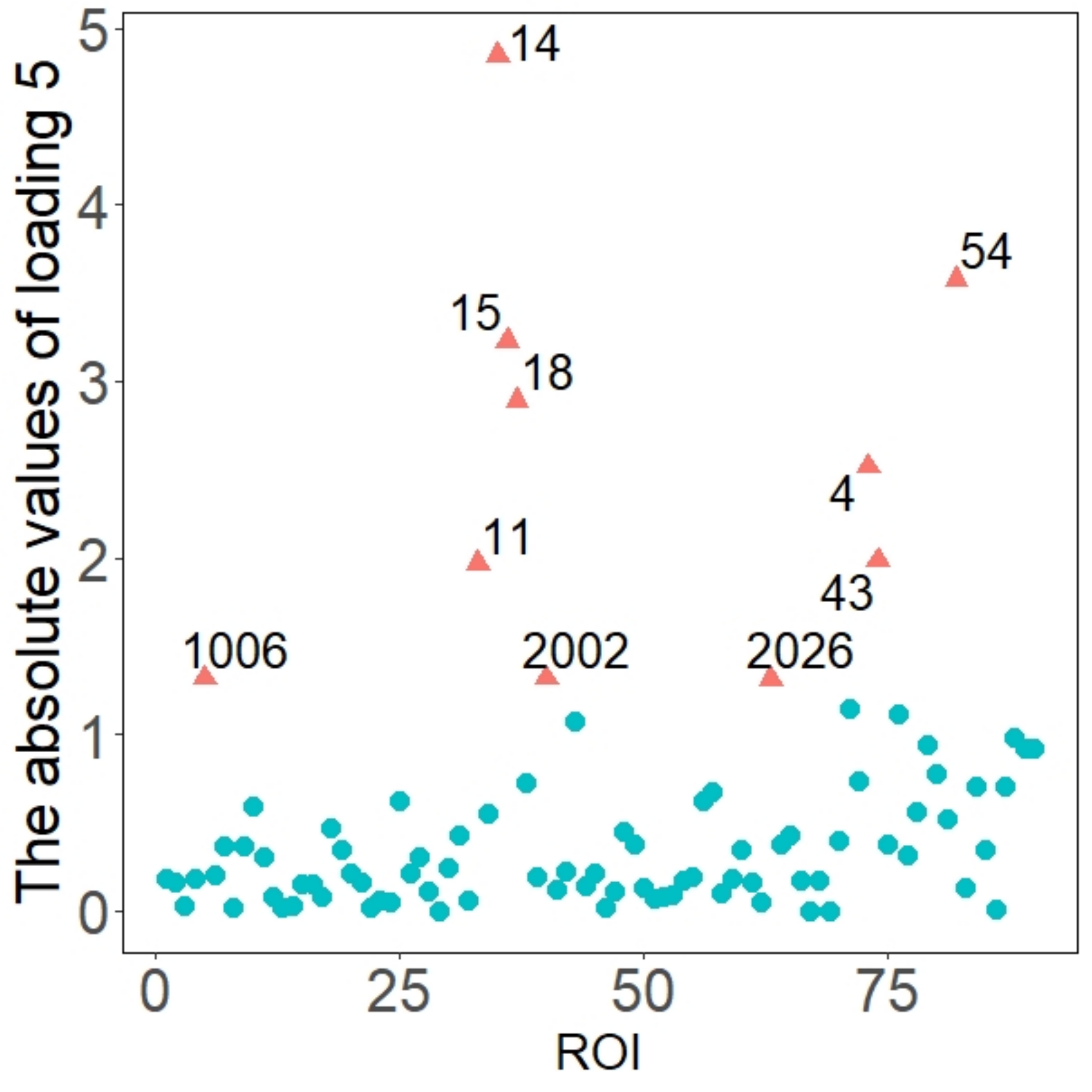}}
\end{center}
		\caption{The absolute values of the 5 loadings, where the red triangular dots are the first 10 largest absolute values of loadings. }
		\label{loading_realdata}
\end{figure}

Since $q=5$ is selected,  we can use the 5-dimensional vector to represent each ROI,
resulting in
$\wh\B\in\RX^{90\times 5}$ for 90 ROIs. Then,  based on the estimated $\wh\B$,  we cluster the ROIs into 3 classes by $K$-means, and the clusters are displayed  in Table \ref{table:cluster}. We can see  that class 1 mainly includes  the ROIs in the basal nucleus, class 2 mainly contains the ROIs  in the  junction between the brain and basal nucleus, and class 3 involves the ROIs in the brain, including  the  lateral and medial parts of the left and right hemispheres.

\begin{table}[!htb]\footnotesize
	\centering
	\caption{The  ROIs and their clusters.}\label{table:cluster}
	\renewcommand\arraystretch{1.2}
	\begin{center}
	\begin{tabular}{llllllllll}
		\hline
		ROI	&	name	&	ROI	&	name	&	ROI	&	name	\\
		\hline
		\textbf{Class 1}\\
		5	&	 left inferior lateral ventricle		&18	&	 left amygdala	&	91	&	 left basal forebrain	\\
		10	&	 left thalamus proper	&	24	&	 CSF	&	92	&	 right basal forebrain	\\
		11	&	 left caudate	&	28	&	 left ventral DC	&	632	&	 cerebellar vermal lobules	\\
		
		12	&	 left putamen	&	49	&	 right thalamus proper	&		&	 	  VIII-X\\
		14	&	 3rd ventricle	&	50	&	 right caudate	&	1014	&	 left medial orbitofrontal	\\
		15	&	 4th ventricle	&	53	&	 right hippocampus	&	1026	&	 left rostral anterior 	\\
		16	&	 Brain stem	&	54	&	 right amygdala	&		&	 cingulate	\\
		17	&	 left hippocampus	&	60	&	 right ventral DC	&	2002	&	 right caudal anterior	\\
		2012	&	 right lateral orbitofrontal	&	2014	&	right medial orbitofronta	&		&	 cingulate	\\
		2023	&right posterior cingulate&&&&
		\\
		\\
		\textbf{Class 2}\\
		51	&	 right putamen	&	1016	&	 left parahippocampal	&	1025	&	 left precuneus	\\
		630	&	 cerebellar vermal lobules I-V	&	1017	&	 left paracentral	&	1035	&	 left insula	\\
		631	&	 cerebellar vermal lobules VI-VII	&	1021	&	 left pericalcarine	&	2013	&	 right lingual	\\
		1012	&	 left lateral orbitofrontal	&	1023	&	 left posterior cingulate	&	2017	&	 right paracentral	\\
		1013	&	 left lingual	&	1024	&	 left precentral	&	2021	&	 right pericalcarine	\\
		2025	&	 right precuneus	&	2026	&	 right rostral anterior cingulate	&		&		\\
		\\
		
		\textbf{Class 3}\\
		4	&	 left lateral ventricle	&	1018	&	 left pars opercularis	&	2010	&	 right isthmus cingulate	\\
		6	&	 left cerebellum exterior	&	1019	&	 left pars orbitalis	&	2011	&	 right lateral occipital	\\
		7	&	 left cerebellum white matter	&	1020	&	 left pars triangularis	&	2015	&	 right mid-temporal	\\
		43	&	 right lateral ventricle	&	1022	&	 left postcentral	&	2016	&	 right parahippocampal	\\
		45	&	 right cerebellum exterior	&	1027	&	 left rostral mid-frontal	&	2018	&	 right pars opercularis	\\
		46	&	 right cerebellum white matter	&	1028	&	 left superior frontal	&	2019	&	 right pars orbitalis	\\
		1002	&	 left caudal anterior cingulate	&	1029	&	 left superior parietal	&	2020	&	 right pars triangularis	\\
		1003	&	 left caudal mid-frontal	&	1030	&	 left superior temporal	&	2022	&	 right postcentral	\\
		1005	&	 left cuneus	&	1031	&	 left supramarginal	&	2024	&	 right precentral	\\
		1006	&	 left entorhinal	&	1034	&	 left transverse temporal	&	2027	&	 right rostral mid-frontal	\\
		1007	&	 left fusiform	&	2003	&	 right caudal mid- frontal	&	2028	&	 right superior frontal	\\
		1008	&	 left inferior parietal	&	2005	&	 right cuneus	&	2029	&	 right superior parietal	\\
		1009	&	 left inferior temporal	&	2006	&	 right entorhinal	&	2030	&	 right superior temporal	\\
		1010	&	 left isthmus cingulate	&	2007	&	 right fusiform	&	2031	&	 right supramarginal	\\
		1011	&	 left lateral occipital	&	2008	&	 right inferior parietal	&	2034	&	 right transverse temporal	\\
		1015	&	 left mid-temporal	&	2009	&	 right inferior temporal	&	2035	&	 right insula	\\
		\hline
	\end{tabular}
\end{center}
\end{table}

We also investigate the effects of 
the ROIs on AD. First, we conduct multivariate logistic regression with AD, MCI and CN as responses and $\wh\bzeta_i$ as covariates. The regression coefficients are denoted by $\ba$.  Hence, $\ba'\wh\bzeta_i$ can be regarded as the measurement 
of the effect of the ROIs on AD.
On the other hand, by multiplying $p^{-1}\PPhi(t)\B'$ on both sides of (\ref{eq:model1})  and coupling it with the identification condition $p^{-1}\B'\B=\I_{q}$, we obtain
$p^{-1}\PPhi(t)\B'\X_i(t)\approx \PPhi(t)\PPhi'(t)\bzeta_i.$
After combining it with the identification condition $\int\PPhi(t)\PPhi(t)' dt=\I$, we have
$$p^{-1}\int\ba' \PPhi(t)\B'\X_i(t) dt \approx \ba'\int\PPhi(t)\PPhi'(t)dt\bzeta_i=\ba'\bzeta_i.$$
Then, the regression relationship  $\ba'\bzeta_i$ between response $Y_i$ and factor $\bzeta_i$ can be written as $\int\bta'(t)\X_i(t)dt$ between $Y_i$ and original functional covariates $\X_i(t)$, where $\bta(t)=(\eta_1(t),\cdots,\eta_p(t))'=p^{-1}\B\PPhi'(t)\ba$ is the regression coefficient function.
To 
visualize the  coefficient function, we choose one ROI from  each  class in Table \ref{table:cluster} and plot the corresponding estimated coefficient function $\wh\eta_j(t)$ for $\eta_j(t)$ in Figure \ref{npex_realdata}(b), which shows that the estimated coefficient function $\wh\bta(t)$  fluctuates violently around a horizontal line. This 
inspires us to  compute the cumulative term $\bb=\int \wh\bta(t) dt$  to measure the effects of the ROIs on the response. Furthermore, for each ROI $j$, we perform hypothesis testing $\beta_j=0$ based on 200 bootstrap samples to test the effect of ROI $j$ on AD. As a result, we identify 41 significant ROIs, as displayed in Table \ref{table:beta}.
All the significant ROI effects are negative,  which  implies  that  brain atrophy is  associated with  AD and is consistent with previous studies \citep{double1996topography}. For example, 
the following conclusions have been found for the 41 significant ROIs:
\begin{itemize}
\item Medial temporal lobe (ROI$_{2015}$) atrophy  was reported by  \cite{fruijtier2019abide}  to be one of the three best validated neuroimaging biomarkers for AD.
\item The tau protein is an indicator of the extent of neuronal damage, and its pathology  was verified by \cite{solders2022diffusion} to  propagate in a prion-like manner along the LC-transentorhinal cortex (ROI$_{1006}$ and ROI$_{2006}$) and white matter (ROI$_{46}$) pathway.
\item In addition,  frontal lobe (ROI$_{1003}$, ROI$_{1028}$,ROI$_{2003}$ and ROI$_{2027}$) atrophy is a prominent symbol of AD and can accelerate the lesions of AD. This was reported by \cite{simona2016chronic}.
\item The atrophy of the parietal lobe (ROI$_{1008}$ and ROI$_{1029}$), insula (ROI$_{2035}$) and hippocampus (ROI$_{17}$ and ROI$_{53}$) was found to induce AD by \cite{foundas1997atrophy}. Moreover, from   Table \ref{table:beta}, we can see  that  the effect of the left hippocampus (ROI$_{17}$) is larger than 
that of the right hippocampus (ROI$_{53}$), which is consistent with the findings of \cite{philip2021alzheimer}, which state that the atrophy of left hippocampus is more severe than 
that of the right hippocampus in AD patients.
\item  The  predictive function of fusiform  gyrus (ROI$_{1007}$) atrophy at the  early stage of AD  has been reported by  \cite{chang2016hippocampal}. \cite{donovan2014regional} indicated damage to the supramarginal gyrus (ROI$_{1031}$) atrophy to the human visual system in AD patients.

\item The associations of 
atrophy in the caudate nucleus (ROI$_{50}$), pars triangularis (ROI$_{1020}$ and ROI$_{2020}$), pars opercularis (ROI$_{1018}$ and ROI$_{2018}$),  and amygdala (ROI$_{54}$)   with AD 
have also been extensively reported in the  literature, for example,  \cite{pearce1984neurotransmitter,whitwell2010progression} and \cite{stephane2011amygdala}.
\end{itemize}

\begin{table}[htb!]
	\centering
	\renewcommand\arraystretch{1.5}\footnotesize
	\caption{Estimates, SDs and $p$ values of $\bb=\int \bta(t) dt$ for the ADNI data, where $\bta(t)$ is the regression coefficient
function of the ROIs.}\label{table:beta}
	\begin{center}
	\begin{tabular}{rrrr|rrrr|rrrr}
		\hline
		ROI	&	Estimate	&	SD	&	$p$ value	&	ROI	&	Estimate	&	SD	&	$p$ value	&	ROI	&	Estimate	&	SD	&	$p$ value	\\
		\hline
		5	&	-5.9829	&	1.9892	&	0.0026	&	1008	&	-1.7920	&	0.6487	&	0.0057	&	2008	&	-2.7590	&	0.7487	&	0.0002	\\
		14	&	-4.0130	&	1.8183	&	0.0273	&	1009	&	-3.8436	&	0.9955	&	0.0001	&	2009	&	-4.8168	&	1.2551	&	0.0001	\\
		17	&	-6.0605	&	1.6149	&	0.0002	&	1011	&	-2.9302	&	0.8575	&	0.0006	&	2010	&	-2.3420	&	0.8745	&	0.0074	\\
		28	&	-9.3520	&	2.4988	&	0.0002	&	1018	&	-3.4751	&	1.0386	&	0.0008	&	2015	&	-3.1155	&	0.8163	&	0.0001	\\
		45	&	-2.5556	&	0.7636	&	0.0008	&	1020	&	-3.6396	&	0.9609	&	0.0002	&	2018	&	-2.4459	&	0.7143	&	0.0006	\\
		46	&	-3.1817	&	1.2945	&	0.0140	&	1028	&	-1.6740	&	0.6912	&	0.0154	&	2019	&	-1.7958	&	0.7866	&	0.0224	\\
		50	&	-7.4244	&	3.1591	&	0.0188	&	1029	&	-3.1541	&	0.8261	&	0.0001	&	2020	&	-1.8009	&	0.6835	&	0.0084	\\
		53	&	-4.9068	&	1.2157	&	0.0001	&	1030	&	-4.1952	&	1.0764	&	0.0001	&	2022	&	-1.2041	&	0.6140	&	0.0499	\\
		54	&	-5.9669	&	1.7803	&	0.0008	&	1031	&	-2.3494	&	0.6523	&	0.0003	&	2027	&	-1.6122	&	0.6545	&	0.0138	\\
		60	&	-10.1306	&	2.8457	&	0.0004	&	1034	&	-3.1273	&	1.0865	&	0.0004	&	2030	&	-3.2471	&	0.8374	&	0.0001	\\
		1003	&	-3.4179	&	0.9363	&	0.0003	&	2003	&	-2.6540	&	0.7662	&	0.0005	&	2031	&	-2.7265	&	0.7388	&	0.0002	\\
		1005	&	-1.8912	&	0.6732	&	0.0049	&	2005	&	-1.7878	&	0.6582	&	0.0066	&	2034	&	-2.0218	&	0.9236	&	0.0286	\\
		1006	&	-6.9455	&	2.1871	&	0.0015	&	2006	&	-2.8513	&	0.8757	&	0.0011	&	2035	&	-1.7238	&	0.7776	&	0.0266	\\
		1007	&	-2.9910	&	0.7921	&	0.0002	&	2007	&	-3.1506	&	0.8217	&	0.0001									\\

		\hline
	\end{tabular}
	\end{center}	
\end{table}

\section{Discussion}

In this paper, we propose 
FaFPCA to  extract   features  from
high-dimensional functional data. By building  some moment equations,  we estimate loadings, scores and eigenfunctions in closed form without rotation.
We also establish the theoretical properties of the proposed estimator, including the  convergence rate  and asymptotic normality. The proposed FaFPCA approach has the following advantages.
(1) Computational efficiency. With the proper rewriting of the FaFPCA model, we directly estimate loadings, scores and eigenfunctions in closed form so that complicated and likely  nonconvergent  iterations can be avoided. 
(2) Blessing of dimensionality.
In high-dimensional FDA, the dimension of function $p$ is a burden for estimation and inference. In contrast, FaFPCA provides functional principal  analysis under the framework of a factor model; as a result, the 
dimensionality curse becomes dimensionality welfare, which is confirmed by Theorems \ref{th1}-\ref{thphi} and simulation studies. 
(3) Theoretical interpretability. We give the theoretical properties of FaFPCA and show the relation to the traditional high-dimensional factor model.
(4) High prediction accuracy. Since we use both the correlation among the high-dimensional functional variables and the dependence over time, the proposed FaFPCA approach has a  higher prediction accuracy than the existing methods, which is  confirmed by real data analysis and simulation studies.

Various extensions can be considered in the future. For example,  we can extend linear FaFPCA to the generalized case for discrete  functional data. However, the  computational  and the theoretical availability  for a nonlinear framework  are worthy of further investigation.
In practice,  the observations $\X_i(t), i = 1, \cdots, n$ may  be serially  dependent, such as those considered by \cite{tavakoli2021high} and \cite{guo2022factor}, it is interesting  to extend FaFPCA to high-dimensional functional time series analysis. In addition,  in the paper, we only consider the dimension reduction of high-dimensional functional data. The features captured by the factors  are the most important directions concerning the information of high-dimensional functional covariates  alone rather than under the supervision of the response. It is an important  question on how to extract factors from high-dimensional functional covariates in the direction supervised by a response variable.

\section*{Acknowledgements}
The functional magnetic resonance imaging (fMRI) scans provided by ADNI database (\url{http://adni.loni.usc.edu/}). ADNI was launched in 2003 by the National Institute on Aging, the National Institute of Biomedical Imaging and Bioengineering, the Food and Drug Administration, private pharmaceutical companies and non-profit organizations as a \$60 million and 5-year public–private partnership. The primary goal of ADNI was to test whether serial MRI, PET and other biological markers are useful in clinical trials of mild cognitive impairment (MCI) and early AD. The determination of sensitive and specific markers of very early AD progression is intended to aid researchers and clinicians to develop new treatments and monitor their effectiveness and estimate the time and cost of clinical trials. ADNI subjects aged 55–90 years old and from over 50 sites across the USA and Canada participated in the research; more detailed information is available at (\url{www.adni-info.org}).

Data used in preparation of this article were obtained from the Alzheimer’s Disease Neuroimaging Initiative (ADNI) database (\url{adni.loni.usc.edu}). As such, the investigators within the ADNI contributed to the design and implementation of ADNI and/or provided data but did not participatein analysis or writing of this report. A complete listing of ADNI investigators can be found at: \url{http://adni.loni.usc.edu/wp-content/uploads/how_to_apply/ADNI_Acknowledgement_List: pdf.}

\bibliographystyle{apalike}
\bibliography{myref}

\end{document}


\def\spacingset#1{\renewcommand{\baselinestretch}%
		{#1}\small\normalsize} \spacingset{1}
	
	\if1\blind{	
		\title{\bf Supplementary Materials to ``Factor-guided functional PCA  for high-dimensional functional data"}
		\author{Shoudao Wen$^1$ and Huazhen Lin$^1$\thanks{
				Corresponding author. Email: linhz@swufe.edu.cn. The research were partially supported by National Natural
				Science Foundation of China (Nos. 11931014 and 11829101).}\hspace{.2cm}\\
			$^1$Center of Statistical Research and  School of Statistics,\\ Southwestern University of
			Finance and Economics}	  
		\maketitle
	}\fi	
	
	\if0\blind{	
		\bigskip
		\bigskip
		\bigskip
		\begin{center}
			{\LARGE\bf Supplementary Materials to ``Factor-guided functional PCA  for high-dimensional functional data"}		 
		\end{center}
		\medskip
	}\fi

\spacingset{1.9}
\linespread{2}
\addtolength{\textheight}{.5in}	
\section{Supplementary Materials A: Notations}
\label{app:notation}

For a matrix $\bm{G}(t)=\{G_{ij}(t);i=1,\cdots, k_1; j=1,\cdots,k_2\}$, define $\int \bm{G}(t)dt=\{\int G_{ij}(t)dt, i=1,\cdots,k_1;j=1,\cdots,k_2\}$. Except for special emphasis,
we omit the integration region $[0,\ 1]$ and omit the dependence of the variable on the subscript $n$ for notation simplicity. To fix notation, let $\|\W\|_1$ be the 1-norm of any matrix $\W$, $\|\W\|_2$ be the spectral-norm, $\|\W\|_F$ be the Frobenius-norm and $\|\W\|_{\infty}$ be the sup-norm.

\section{Supplementary Materials B: Proof}\label{app:proof}
\textbf{Proof of Proposition 1:} Denote $\X(t)=(\X_1(t),\cdots,\X_n(t))$ and $\u(t)=(\u_{1}(t),\cdots,\u_{n}(t))$, then model (\ref{eq:model1}) can be written as $\X(t)=\B\PPhi'(t)\bzeta+\u(t)$.

a) We first prove $\bzeta$ is identifiable. We note that
\begin{equation}\label{xiide1}
\frac{1}{np}\int\X'(t)\X(t)dt=\frac{1}{n}\bzeta'\bzeta+\frac{1}{np}\bzeta'\int\PPhi(t)\B'\u(t)dt+\frac{1}{np}\int\u'(t)\B\PPhi'(t)dt\bzeta+\frac{1}{np}\int\u'(t)\u(t) dt,
\end{equation}
and
\begin{equation}\label{xiide2}
\|\frac{1}{np}\int\u'(t)\B\PPhi'(t)dt\bzeta\|_F\leq\left(\frac{1}{n}\|\bzeta\|^2_2\cdot \big(\frac{1}{np}\sumi\sum_{k=1}^{q}\sum_{k'=1}^K \int(\sumj b_{jk}u_{ij}(t)\phi_{kk'}(t))^2dt\big)\cdot\frac{1}{p} \right)^{\frac{1}{2}}
=o_p(1).
\end{equation}
We consider $\sumj u_{ij}(t)u_{i'j}(t)$ for $i\neq i'$ and given $t$. First, we have $\Ex\big(\sumj u_{ij}(t)u_{i'j}(t)\big)=0$.
Then we calculate the variance as
\begin{equation*}
\var\big(\sumj u_{ij}(t)u_{i'j}(t)\big) = \Ex\big(\sumj u_{ij}(t)u_{i'j}(t)\big)^2
\leq\sumj \big(\sum_{j'=1}^p\big| \Ex(u_{ij}(t)u_{ij'}(t)) \big|\big)^2=O_p(p)
\end{equation*}
by Condition (A3), thus it follows that $n^{-2}p^{-2}\sumi\sum_{i'\neq i}\big(\sumj u_{ij}(t)u_{i'j}(t)\big)^2=O_p(p^{-1}).$
Further, we note that
\begin{equation*}
\frac{1}{n^2p^2}\sumi\big(\sumj u^2_{ij}(t)\big)^2=O_p\Big\{
\frac{1}{n^2p^2}\sumi\Big(\sumj \big[u^2_{ij}(t)-\Ex(u^2_{ij}(t))\big]\Big)^2
+ \frac{1}{n^2p^2}\sumi\big( \sumj\Ex(u^2_{ij}(t))\big)^2
\Big\}.
\end{equation*}
For the first term,  similar to the proof of part $(b)$ in Lemma A.2 in \cite{bai2003inferential}, we have
$
n^{-1}p^{-2}\sum_{i'=1}^n\big(\sumj \big[u^2_{ij}(t)-\Ex(u^2_{ij}(t)\big] \big)^2=O_p(p^{-1})
$
by Condition (A3),
thus the first term is $O_p(p^{-1})$ and it easy to check that the second term is $O_p(n^{-1})$.
Thus, we have $\|(np)^{-1}\int\u'(t)\u(t)dt\|_F=O_p( p^{-\frac{1}{2}})=o_p(1)$.
Then (\ref{xiide1}) and (\ref{xiide2}) yield
\begin{equation}\label{eq:zeta1}
\frac{1}{n}\|\frac{1}{p}\int\X'(t)\X(t)dt-\bzeta'\bzeta\|_2\leq\|\frac{1}{np}\int\X'(t)\X(t)dt-\frac{1}{n}\bzeta'\bzeta\|_F=o_p(1).
\end{equation}
Let $\S\bm{\psi}^2\S'$ be the SVD of $p^{-1}\int\X'(t)\X(t)dt$, where $\S=(\s_1,\cdots,\s_n)$  and the first nonzero element of  $\s_l$ is positive for $l=1,\cdots,n$ and $\bm{\psi}^2=\diag(\psi^{*2}_1,\cdots,\psi^{*2}_n)$ with $\psi^{*2}_1\geq\cdots\geq\psi^{*2}_n\geq 0$. We further define $\S_{Kq}=(\s_1,\cdots,\s_{Kq})$ and $\bm{\psi}^2_{Kq}=\diag(\psi^{*2}_1,\cdots,\psi^{*2}_{Kq})$.
Letting $\C\Psi\L'$ be the SVD of $\bzeta'$, we have $\bzeta'\bzeta=\C\Psi\L'\L\Psi\C'=\C\Psi^2\C'$, which implies $\C\Psi^2\C'$ is the eigenvalue decomposition of $\bzeta'\bzeta$.
Following the same line in  \cite{asz010}, we have $\L$ is a sign matrix (diagonal matrix with 1 and $-1$ on the diagonals)
by Conditions (A1) and (A3) and (\ref{eq:zeta1}). Thus, we can get that as $n,p,K\to\infty$,
\begin{equation*}
\bm{\psi}_{Kq}^2,\S_{Kq},\Psi,\C\ \text{can be identified and} \ n^{-\frac{1}{2}}\|\Psi-\bm{\psi}_{Kq}\|_2\to0,   \|\C-\S_{Kq}\|_2\to 0.
\end{equation*}
Since $\bzeta'=\C\Psi\L'$, by Condition (A1) and ($\ref{xiide1}$), we have
\begin{equation*}
\begin{split}
n^{-\frac{1}{2}}\|\bzeta'-\S_{Kq}&\bm{\psi}_{Kq}\L'\|_2
\leq\|\C n^{-\frac{1}{2}}(\Psi-\bm{\psi}_{Kq})\L'\|_2 + \|(\C-\S_{q})n^{-\frac{1}{2}}\bm{\psi}_{Kq}\L'\|_2
\to0.
\end{split}
\end{equation*}
Noting that the first nonzero element in each column of $\S_{Kq}$ is positive, $\|\C-\S_{Kq}\|_2\to 0$ implies that
the first non-zero element in each column of $\C$  is also positive
when $n$ is sufficiently large. Hence by (I2), we conclude that $\L$ is an identity matrix. Thus, $n^{-\frac{1}{2}}\|\bzeta'-\S_{Kq}\bm{\psi}_{Kq}\|_2\to0$, and then $\bzeta$ is unique.

b) We then show $\B$ is unique following the similar line to the proof of $\bzeta$.
Denote $\bSig_{\bzeta}$, $\bSig_{\X}(t)$ and $\bSig_{\u}(t)$ be the covariance matrix of $\bzeta_i,$
$\X_i(t)$ and $\u_i(t)$ respectively for the fixed $t$. Further, denote $\wt\bSig_{\X}=\int\bSig_{\X}(t) dt$ and $\wt\bSig_{\u}=\int\bSig_{\u}(t) dt$. Then
\begin{equation*}
\label{eq1}
\wt\bSig_{\X}=\int\B\PPhi'(t)\bSig_{\bzeta}\PPhi(t)\B' dt + \wt\bSig_{\u}=\B\LLambda_{\bzeta}\B'+\wt\bSig_{\u},
\end{equation*}
where $\LLambda_{\bzeta}=\diag\big(\sumk\var(\xi_{i1k}),\cdots,\sumk\var(\xi_{iqk})\big)$.
Since $\|\LLambda_{\bzeta}\|_2=O_p(1)$ by Condition (A1) and  $\|\wt\bSig_{\u}\|_1=\max_{1\leq j\leq p}\sum_{j'=1}^p|\Ex(\int u_{ij}(t)u_{ij'}(t)dt) | \leq C<\infty$ by Condition (A3), we obtain $p^{-1}\|\wt\bSig_{\X}-\B\LLambda_{\bzeta}\B'\|_1=p^{-1}\|\wt\bSig_{\u}\|_1\to 0.$

Let $\W\R^2\W'$ be the SVD of $\wt\bSig_{\X}$.
where $\W=(\w_1,\cdots,\w_n)$  and the first nonzero element of  $\w_l$ is positive for $l=1,\cdots,p$ and $\bm{\R}^2=\diag(r^{*2}_1,\cdots,r^{*2}_n)$ with $r^{*2}_1\geq\cdots\geq r^{*2}_n\geq 0$. We further define $\W_{q}=(\w_1,\cdots,\w_{q})$ and $\bm{\R}^2_{q}=\diag(r^{*2}_1,\cdots,r^{*2}_{q})$.
Letting $\Q\D\K'$ be the SVD of $\wt\B=\B\LLambda_{\zeta}^{\frac{1}{2}}$,
$\LLambda_{\zeta}^{\frac{1}{2}}$ can be well defined and $\LLambda_{\zeta}^{-\frac{1}{2}}$ exists because $\LLambda_{\zeta}$ is diagonal.
Then $\B=\Q\D\K'\LLambda_{\zeta}^{-\frac{1}{2}}$ and $\B\LLambda_{\bzeta}\B'=\Q\D\K'\K\D\Q'=\Q\D^2\Q',$ which implies $\Q\D^2\Q'$ is the eigenvalue decomposition of $\B\LLambda_{\bzeta}\B'$. Thus,
\begin{eqnarray}\label{eq2}
p^{-1}\|\W\R^2\W'-\Q\D^2\Q' \|_2&=p^{-1}\|\wt\bSig_{\X}-\B\LLambda_{\bzeta}\B'\|_2\leq p^{-1}\sqrt{\|\wt\bSig_{\X}-\B\LLambda_{\bzeta}\B'\|_1\|\wt\bSig_{\X}-\B\LLambda_{\bzeta}\B'\|_{\infty} }\nonumber\\
&=p^{-1}\|\wt\bSig_{\X}-\B\LLambda_{\bzeta}\B' \|_1\to 0.
\end{eqnarray}
Following the same line in  \cite{asz010}, we have by (\ref{eq2}), Conditions (A1) and (A3) that
\begin{equation}\label{eq3}
\R_{q}^2,\W_{q},\D,\Q\ \text{can be identified and} \ \|\Q-\W_{q}\|_2\to0,   p^{-\frac{1}{2}}\|\D-\R_{q}\|_2\to 0.
\end{equation}
Noting $\wt\B=\Q\D\K'$,  we have by Condition (A1) and ($\ref{eq3}$) that
\begin{equation}\label{eq4}
\begin{split}
p^{-\frac{1}{2}}\|\wt\B-\W_{q}&\R_{q}\K'\|_2\leq\|\Q p^{-\frac{1}{2}}(\D-\R_{q})\K'\|_2+\|(\Q-\W_{q})p^{-\frac{1}{2}}\R_{q}\K'\|_2
\to0.
\end{split}
\end{equation}
Thus, $\K$ is an identity matrix, and $p^{-\frac{1}{2}}\|\B\LLambda_{\bzeta}^{\frac{1}{2}}-\W_{q}\R_{q}\|_2\to0$.
Then $p^{-\frac{1}{2}}\|\B-\W_{q}\R_{q}\LLambda_{\bzeta}^{-\frac{1}{2}}\|_2\leq p^{-\frac{1}{2}}\|\B\LLambda_{\bzeta}^{\frac{1}{2}}-\W_{q}\R_{q}\|_2\|\LLambda_{\bzeta}^{-\frac{1}{2}}\|_2=o_p(1).$
Since $\bzeta$ is unique, it follows that $\LLambda_{\bzeta}$ is unique, and then we get that $\B$ is unique.

c) We show $\h_i(t)$ is identifiable. Model (\ref{eq:model10}) gives
\begin{equation}\label{eq5}
\begin{split}
\|p^{-\frac{1}{2}}\X_i(t) \|_2^2
\leq 2\int\bzeta_i'\PPhi(t)\PPhi'(t)\bzeta_i dt+2p^{-1}\int\u_i'(t)\u_i(t)dt=2\bzeta_i'\bzeta_i+2p^{-1}\int\u_i'(t)\u_i(t)dt=O_p(1),
\end{split}
\end{equation}
where the last equality holds because $\bzeta_i'\bzeta_i=O_p(1)$ by Condition (A2) and $p^{-1}\int \u_i'(t)\u_i(t) dt=O_p(1)$.

Thus, by (\ref{eq4}) and (\ref{eq5}) we have
\begin{equation*}
\begin{split}
\|p^{-1}\B'\X_i(t)-p^{-1}\LLambda^{-\frac{1}{2}}_{\bzeta}\R_{q}'\W_{q}\X_i(t)\|_2\leq p^{-\frac{1}{2}}\|\B-\LLambda^{-\frac{1}{2}}_{\bzeta}\W_{q}\R_{q}\|_2\|p^{-\frac{1}{2}}\X_i(t)\|_2\to0.
\end{split}
\end{equation*}
Therefore,
\begin{equation*}
\lim_{p\to\infty}p^{-1}\LLambda^{-\frac{1}{2}}_{\bzeta}\R_{q}'\W_{q}\X_i(t) =
\lim_{p\to\infty} p^{-1}\B'\X_i(t)-p^{-1}\B'\u_i(t) = \h_i(t),
\end{equation*}
almots surely. Since $\R_{q}, \W_{q}$ are unique, it follows $\h_i(t)$ is unique.

d) We show $\PPhi(t)$ is unique. Noting $\X(t)=\B\PPhi'(t)\bzeta+\u(t)$, we have

\begin{equation}\label{eq:phi1}
\begin{split}
\lim_{n,p\to\infty}(np)^{-1}\bSig^{-1}_{\bzeta}\bzeta\X'(t)\B&-\lim_{n,p\to\infty}(np)^{-1}\bSig^{-1}_{\bzeta}\bzeta\u'(t)\B
=\PPhi(t).
\end{split}
\end{equation}

By Conditions (A1) and (A4), we have
\begin{equation}\label{eq:phi2}
\|\frac{1}{np}\bSig^{-1}_{\bzeta}\bzeta\u'(t)\B\|_F\leq\left(\|\bSig^{-1}_{\bzeta}\|^2_2\cdot\frac{1}{n}\|\bzeta\|^2_2\cdot \big(\frac{1}{np}\sumi\sumj\sum_{k=1}^{q}\int(b_{jk}u_{ij}(t))^2dt\big)\cdot\frac{1}{p} \right)^{\frac{1}{2}}=o_p(1).
\end{equation}
(\ref{eq:phi1}) and (\ref{eq:phi2}) imply
$\PPhi(t)=\lim_{n,p\to\infty}(np)^{-1}\bSig^{-1}_{\bzeta}\bzeta\X'(t)\B$ almost surely.
Because $\B$ and $\bzeta$ are unique,  $\PPhi(t)$ is unique.
\\
\\

\textbf{Proof of Theorem 1:}
We first consider the convergence rate of $\|\wh\b_j-\b_{j0}\|_2.$ Denote $\X_i(\t_i)=(\X_{i1}(\t_i),\cdots,\X_{ip}(\t_i))\in\RX^{p\times n_i}$ with $\X_{ij}(\t_i)=(X_{ij}(t_{i1}),\cdots,X_{ij}(t_{i,n_i}))'.$
Letting $\wc V$ be the $q\times q$ diagonal matrix composed by the first $q$ largest
eigenvalues of $(np)^{-1}\sumi n^{-1}_i\X_i(\t_i)\X'_i(\t_i)$ in decreasing order.
We have $\wh\B=(np)^{-1}\sumi n^{-1}_i\X_i(\t_i)\X'_i(\t_i)\wh\B\wh V^{-1}$.
Let $\wc H
=p^{-1}\wc V^{-1}\wh\B'\B_0\LLambda_{\bzeta}$ be a $q\times q$ matrix. Thus,
\begin{equation}\label{eq:wcb1}
\begin{split}
\wh\b_j-\wc H\b_{j0}&=\frac{1}{np}\wc V^{-1}\sum_{j'=1}^p\wh\b_{j'}\b'_{j'0}\sumi\sum_{l=1}^{n_i}\frac{1}{n_i}\PPhi_0(t_{il})
\bzeta_{i0}u_{ij}(t_{il})
+\frac{1}{np}\wc V^{-1}\sum_{j'=1}^p\wh\b_{j'}
\sumi\sum_{l=1}^{n_i}\frac{1}{n_i}u_{ij'}(t_{il})\bzeta'_{i0}\PPhi_0(t_{il})\b_{j0}
\\&\ \ \ \ \ \ \ \ +\frac{1}{np}\wc V^{-1}\sum_{j'=1}^p\wh\b_{j'}
\sumi\sum_{l=1}^{n_i}\frac{1}{n_i}u_{ij}(t_{il})u_{ij'}(t_{il})
+O_p(N_0^{-1/2})\\
&\triangleq\Rmnum{1}_b+\Rmnum{2}_b+\Rmnum{3}_b+O_p(N_0^{-1/2}),
\end{split}
\end{equation}
where the last term $O_p(N_0^{-1/2})$ is the numerical integral approximation error by noting $\Ex\Big(\frac{1}{n}\sumi\frac{1}{n_i}\sum_{l=1}^{n_i}\big(h^2_{ij}(t_{il})-\Ex(\int h^2_{ij}(t) dt)
\big)\Big)=0$ and $\var\Big(\frac{1}{n}\sumi\frac{1}{n_i}\sum_{l=1}^{n_i}\big(h^2_{ij}(t_{il})-\Ex(\int h^2_{ij}(t) dt)\Big)=O_p(\frac{1}{n^2}\sumi\frac{1}{n_i}).$

For $\Rmnum{1}_b$, we have $\|\Rmnum{1}_b\|_2\leq
(np)^{-1}\|\wc V^{-1}\|_2 \cdot \|\wh\B\|_2 \cdot \|\B_0\|_2 \cdot \|\sumi\sum_{l=1}^{n_i}\frac{1}{n_i}\PPhi_0(t_{il})
\bzeta_{i0}u_{ij}(t_{il})\|_F,$
and $\|\wh\B\|_2=\sqrt{\lambda_{\max}(\wh\B'\wh\B)}=\sqrt{\lambda_{\max}(p\I_{q})}=O_p(p^{\frac{1}{2}}).$
We first show that $\|\wc V^{-1}\|_2=O_p(1)$. Using $(\ref{eq2})$, by the definition of $\wc V$, we have $\wc V=p^{-1}\R_{q}^2+o_p(1)$ and $\|\wc V-\LLambda_{\bzeta}\|_2=o_p(1).$
By Lemma 2 of \cite{asz010}, we have $\|\wc V^{-1}-\LLambda_{\bzeta}^{-1}\|_2
=O_p(1)\cdot\|\wc V-\LLambda_{\bzeta}\|_2
=o_p(1)$. It follows $\|\wc V^{-1}\|_2
=\|\wc V^{-1}-\LLambda_{\bzeta}^{-1} + \LLambda_{\bzeta}^{-1}\|_2=O_p(1).$
Moreover, note that
\begin{equation}\label{eq:bchek1}
\begin{split}
\sum _{l=1}^{q}\sumi \sumk \xi^2_{ilk,0}
(\frac{1}{n_i}\sum_{l'=1}^{n_i}\phi^2_{lk}(t_{il'}) )
(\frac{1}{n_i}\sum_{l'=1}^{n_i}u^2_{ij}(t_{il'}) )=O_p(n).
\end{split}
\end{equation}
Besides, $\Ex(\eta)=0$  and $\var(\eta)=\Ex(\eta^2)=O_p(n^2)$ with $\eta=\sum\limits_{l=1}^{q}\sum\limits_{i\neq i'}\sum\limits_{k\neq k'}\xi_{ilk,0}\xi_{i'lk',0}\big(\frac{1}{n_i}\sum_{l'=1}^{n_i}\phi_{lk}(t_{il'})\\u_{ij}(t_{il'})\big)
\big(\frac{1}{n_{i'}}\sum_{l'=1}^{n_{i'}}\phi_{lk'}(t_{i'l'})u_{i'j}(t_{i'l'})\big)
$,
since $\Ex(\xi_{ilk,0}\xi_{i'l'k',0})=0$ for $i=i', l=l', k=k', \Ex(u_{ij}(t))=0$ and $\Ex(\xi_{ilk,0}u_{ij}(t))=0$, which means $\eta=O_p(n).$
Combining (\ref{eq:bchek1}) yields $\|\sumi\sum_{l=1}^{n_i}\frac{1}{n_i}\PPhi_0(t_{il})
\\\bzeta_{i0}u_{ij}(t_{il})\|^2_F =O_p(n^{\frac{1}{2}})$.
This gives $\|\Rmnum{1}_b\|_2=
O_p(n^{-\frac{1}{2}})$.	

We then split $\Rmnum{2}_b$ into
$(np)^{-1}\wc V^{-1}\sum _{j'=1}^p(\wh\b_{j'}
-\wc H\b_{j'0})\big(\sumi\sum_{l=1}^{n_i}\frac{1}{n_i}u_{ij'}(t_{il})\bzeta'_{i0}\PPhi_0(t_{il})\big)\b_{j0}
+(np)^{-1}\wc V^{-1}\wc H\sum _{j'=1}^p\b_{j'0}\big(\sumi\sum_{l=1}^{n_i}\frac{1}{n_i}u_{ij'}(t_{il})\bzeta'_{i0}\PPhi_0(t_{il})\big)\b_{j0}.$
Similar to the proof of (2) in \cite{bai2013principal}, we have $p^{-1}\wh\B'\B_0=p^{-1}(\wh\B'-\wc H\B_0'+\wh H\B_0')\B_0=\wc H+O_p(n^{-1}+p^{-1})$
and
$p^{-1}\wh\B\B_0\wc H'=p^{-1}\wh\B(\B_0\wc H'-\wh\B+\wh\B)=\I_{q}+O_p(n^{-1}+p^{-1})$.
Thus, $\I_{q}=\wc H\wc H'+O_p(n^{-1}+p^{-1}).$
This shows that $\wc H'$ is an orthogonal matrix so that its eigenvalues are either 1 or $-1$ up to the order of $O(n^{-1}+p^{-1})$. The definition of $\wc H$ yields $\wc V\wc H=\wc H\LLambda_{\bzeta}$,
which implies that $\wc H$ is diagonal
consisting of the eigenvectors of the diagonal matrix $\LLambda_{\bzeta}$ up to the order of $O(n^{-1}+p^{-1})$.
It follows that
\begin{eqnarray}\label{wcH}
\|\wc H-\I_{q}\|_F=O_p(n^{-1}+p^{-1}).
\end{eqnarray}
Recalling $\wh\B=(np)^{-1}\sumi n^{-1}_i\X_i(\t_i)\X'_i(\t_i)\wh\B\wh V^{-1},$
we can get that the first term of $\Rmnum{2}_b$ is of the order $O_p(n^{-1}+p^{-1})$ and the second term is of the order $O_p(n^{-\frac{1}{2}}p^{-\frac{1}{2}})$ 
by noting that $\|\wc H-\I_{q}\|_F=O_p(n^{-1}+p^{-1})$.
Thus, we have $\|\Rmnum{2}_b\|_2=O_p(n^{-1} + p^{-1} + n^{-\frac{1}{2}}p^{-\frac{1}{2}}).$

At last, we consider $\Rmnum{3}_b=(np)^{-1}\wc V^{-1}\sum _{j'=1}^p\wh\b_{j'}\sumi\sum_{l=1}^{n_i}\frac{1}{n_i}[u_{ij}(t_{il})u_{ij'}(t_{il})-\Ex(u_{ij}(t_{il})u_{ij'}(t_{il}))]
+(np)^{-1}\wc V^{-1}\sum _{j'=1}^p\wh\b_{j'}\sumi\sum_{l=1}^{n_i}\frac{1}{n_i}\Ex(u_{ij}(t_{il})u_{ij'}(t_{il})).$
Similar to the proof of part $(b)$ in Lemma A.2 in \cite{bai2003inferential}, the first term is $O_p(n^{-1} + n^{-\frac{1}{2}}p^{-\frac{1}{2}})$. Then we have
\begin{equation*}
\|\frac{1}{np}\wc V^{-1}\sum _{j'=1}^p\wh\b_{j'}\sumi\sum_{l=1}^{n_i}\frac{1}{n_i}\Ex(u_{ij}(t_{il})u_{ij'}(t_{il}))\|_2\leq\frac{1}{np}\|\wc V^{-1}\|_2\cdot\|\wh\B\|_2\cdot\Big(
\sumi\sumj\big|\Ex(u_{{ij}}(t)u_{ij'}(t))\big|
\Big)=O_p(p^{-\frac{1}{2}})
\end{equation*}
by Conditions (A3) and (I1). And $\|\Rmnum{3}_b\|_2=O_p(n^{-1} + n^{-\frac{1}{2}}p^{-\frac{1}{2}} +p^{-\frac{1}{2}})$.
Therefore, $\|\wh\b_j-\wc H\b_{j0}\|_2=O_p(n^{-\frac{1}{2}}+p^{-\frac{1}{2}} + N_0^{-1/2})$ by (\ref{eq:wcb1}).
Hence, $\|\wh\b_j-\b_{j0}\|_2=\|\wh\b_j-\wc H\b_{j0}\|_2+\|\wc H-\I_{q}\|_F\|\b_{j0}\|_2=O_p(n^{-\frac{1}{2}}+p^{-\frac{1}{2}}
+N_0^{-1/2})$ by (\ref{wcH}).

To get the uniform rate of $\wh\b_j$, we first have $\mathop{\sup}_j\|\wh\b_j-\b_{j0}\|_2=O_p(N^{-1/2}_0+n^{-1}+p^{-1}+n^{-\frac{1}{2}}p^{-\frac{1}{2}})+\mathop{\sup}_j\|\Rmnum{1}_b\|_2+\mathop{\sup}_j\|\Rmnum{3}_b\|_2$ by noting $\mathop{\sup}_j\|\b_{j0}\|=O_p(1)$ by Condition (A2). So the rate of numerical integral approximation error and $\Rmnum{2}_b$ will not change, even we take supremum of $j$. Then following the similar proof of  A.5 in \cite{bai2013statistical}, we have $\mathop{\sup}_j\|\Rmnum{1}_b\|_2=O_p((\log p)^{1/2}n^{-1/2}),$
$\mathop{\sup}_j\|\Rmnum{3}_b\|_2=O_p((\log p)^{1/2}n^{-1/2}+p^{-1/2}).$ Thus, we have $\mathop{\sup}_j\|\wh\b_j-\b_{j0}\|_2=O_p(N^{-1/2}_0+(\log p)^{1/2}n^{-1/2}+p^{-1/2}+n^{-1}+p^{-1}+n^{-\frac{1}{2}}p^{-\frac{1}{2}})=O_p((\log p)^{1/2}n^{-1/2}+p^{-1/2}).$ This completes the proof of Theorem 1.
\\
\\

\textbf{Proof of Theorem 2:}
Now we are at the position to show the convergence rate of $\wh\bzeta_i$.
Denote $\M^{*}(t)=\I_{q}\otimes\M(t)$, where $\otimes$ is Kronecker product and $\K_i=\sum_{l=1}^{n_i}\M^{*}(t_{il})\M^{*\prime}(t_{il})$,
we have $\wh\bzeta_i=n^{-1}p^{-2}\tau_n^{-1}\wt V^{-1}
\Big(\sum_{j=1}^n\wh\bzeta_j\big(\sum_{l=1}^{n_j}\X'_{j}(t_{jl})\wh\B\M^{*\prime}(t_{jl})\big)\K^{-1}_j\Big)\K^{-1}_i\big(
\sum_{l=1}^{n_i}\M^{*}(t_{il})\wh\B'\X_i(t_{il})
\big)$, where $\wt V=n^{-1}\wh\bzeta\wh\bzeta'.$ We note that
\begin{equation}\label{eq:zetab1}
\begin{split}
\wh\bzeta_i-n^{-1}&p^{-2}\tau_n^{-1}\wt V^{-1}
\Big(\sum_{j=1}^n\wh\bzeta_j\big(\sum_{l=1}^{n_j}\X'_{j}(t_{jl})\B_0\M^{*\prime}(t_{jl})\big)\K^{-1}_j\Big)\K^{-1}_i\big(
\sum_{l=1}^{n_i}\M^{*}(t_{il})\B'_0\X_i(t_{il})
\big)\\
=n^{-1}&p^{-2}\tau_n^{-1}\wt V^{-1}
\Big(\sum_{j=1}^n\wh\bzeta_j\big(\sum_{l=1}^{n_j}\X'_{j}(t_{jl})(\wh\B-\B_0)\M^{*\prime}(t_{jl})\big)\K^{-1}_j\Big)\K^{-1}_i\big(
\sum_{l=1}^{n_i}\M^{*}(t_{il})\B'_0\X_i(t_{il})
\big)\\
+& n^{-1}p^{-2}\tau_n^{-1}\wt V^{-1}
\Big(\sum_{j=1}^n\wh\bzeta_j\big(\sum_{l=1}^{n_j}\X'_{j}(t_{jl})\B_0\M^{*\prime}(t_{jl})\big)\K^{-1}_j\Big)\K^{-1}_i\big(
\sum_{l=1}^{n_i}\M^{*}(t_{il})(\wh\B'-\B'_0)\X_i(t_{il})
\big)\\
+& n^{-1}p^{-2}\tau_n^{-1}\wt V^{-1}
\Big(\sum_{j=1}^n\wh\bzeta_j\big(\sum_{l=1}^{n_j}\X'_{j}(t_{jl})(\wh\B-\B_0)\M^{*\prime}(t_{jl})\big)\K^{-1}_j\Big)\K^{-1}_i\big(
\sum_{l=1}^{n_i}\M^{*}(t_{il})(\wh\B'-\B'_0)\X_i(t_{il})
\big)\\
\triangleq\ \Rmnum{1}_{b\zeta}&+\Rmnum{2}_{b\zeta}+\Rmnum{3}_{b\zeta}
\end{split}
\end{equation}

We first give the convergence rate of $\wh\bzeta_i.$
First, we consider $\Rmnum{1}_{b\zeta}$. Note
\begin{equation}\label{eq:zetab2}
\|\Rmnum{1}_{b\zeta}\|_2\leq\frac{1}{np^2\tau_n}\|\wt V^{-1}\|_2\cdot \|\wh\bzeta\|_F\cdot\Big\|
\big(\sum_{j=1}^n\sum_{l=1}^{n_j}\X'_{j}(t_{jl})(\wh\B-\B_0)\M^{*\prime}(t_{jl})\K^{-1}_j\big)\K^{-1}_i\big(
\sum_{l=1}^{n_i}\M^{*}(t_{il})\B'_0\X_i(t_{il})
\big)
\Big\|_F.
\end{equation}
Then we have $\|\Rmnum{1}_{b\zeta}\|_2=O_p(n^{-\frac{1-e-v}{2}}+n^{\frac{e+v}{2}}p^{-\frac{1}{2}}+n^{\frac{e+v}{2}}N_0^{-1/2}),$
by noting $\|\wh\bzeta\|_F=O_p(K^{1/2}\cdot\|\wh\bzeta\|_2)=O_p(n^{\frac{1+e}{2}})$ and $\|\wt V^{-1}\|_2=O_p(1)$.

In the following part, we will give the proof of  $\|\wh\bzeta\|_2=O_p(n^{\frac{1}{2}})$ and $\|\wt V^{-1}\|_2=O_p(1)$.
Denote $\L^{*}=(\L^{*}_{ij})\in\RX^{n\times n}$ with $\L^{*}_{ij}=
n^{-1}p^{-2}K\tau_n^{-1}
\big(\sum_{l=1}^{n_i}\X'_i(t_{il})\B\M^{*\prime}(t_{il})\big)
\K^{-1}_i\K^{-1}_j\big(
\sum_{l=1}^{n_j}\M^{*}(t_{jl})\\\B'\X_j(t_{jl})
\big).
$ Multiplying both sides of
$\wh\bzeta=\wt V^{-1}\wh\bzeta\wh\L^{*}$ by
$n^{-\frac{1}{2}}\wt V^{\frac{1}{2}}$,
we have $ \wt V(n^{-\frac{1}{2}}\wt V^{-\frac{1}{2}}\wh\bzeta)
= n^{-\frac{1}{2}}\wt V^{\frac{1}{2}}\wh\bzeta  =
n^{-\frac{1}{2}}\wt V^{-\frac{1}{2}}\wh\bzeta\L^{*}.$
Because $(n^{-\frac{1}{2}}\wt V^{-\frac{1}{2}}\wh\bzeta)  (n^{-\frac{1}{2}}\wt V^{-\frac{1}{2}}\wh\bzeta)'=\I_{Kq}$ and $\wt V$ is diagonal matrix with decreasing entries, $\wt V$ is the $Kq\times Kq$ diagonal matrix consisting of the first $Kq$ largest eigenvalues of $\wh\L^{*}$ and  $n^{-\frac{1}{2}}\wt V^{-\frac{1}{2}}\wh\bzeta$ is the corresponding eigenvector matrix. 
Note
\begin{equation}\label{eq:et}
\X_i(t)=\B_0\PPhi_0'(t)\bzeta_i+\u_i(t)=\B_0\M^{*\prime}(t)\bTheta'_0\bzeta_i+\B_0\e'(t)\bzeta_i+\u_i(t),
\end{equation}
where $\e(t)=\PPhi_0(t)-\PPhi_{n0}(t)=\diag(\e_1(t),\cdots,\e_{q}(t))$ is a $Kq\times q$ block diagonal matrix with block $j$ being $\e_j(t)=(e_{j1}(t),\cdots,e_{jK}(t))'$
and $e_{jk}(t)=\phi_{jk,0}(t)-\phi_{jk,n0}(t).$
By the asymptotic property of spline approximation, we have $\mathop{\sup} _{t\in[0,\ 1]} |e_{jk}(t)|= \mathop{\sup} _{t\in[0,\ 1]}|\phi_{jk,n0}(t)-\phi_{jk,0}(t)|=O_p(n^{-rv})$ for each $j=1,\cdots,q, k=1,\cdots,K.$
Thus, $\|\wh\L^{*}-\frac{1}{n}\bzeta'_0\bzeta_0 \|_2\leq\|\wh\L^{*}-\L^{*}_0\|_2+\|\L^{*}_0- \frac{1}{n}\bzeta'_0\bzeta_0\|_2
=o_p(1).$
In the last equality of the above equation,
the first term 
is concluded similar to $(\ref{eq:zetab1})$ and the second term is similar to (\ref{eq:zeta_2}) based on (\ref{eq:et}).
Because the first $Kq$ largest eigenvalues of $n^{-1}\bzeta'_0\bzeta_0$ equal to those of $
n^{-1}\bzeta_0\bzeta'_0$ and $n^{-1}\bzeta_0\bzeta'_0$ is diagonal,  $n^{-1}\bzeta_0\bzeta'_0$ is the diagonal matrix with decreasing diagonal entries which consists of the first $Kq$ largest eigenvalues of $n^{-1}\bzeta'_0\bzeta_0$.
Because $\wt V$  is the diagonal matrix with decreasing diagonal entries which consists of the first $Kq$ largest eigenvalues of $\wh\L^{*}$, we then have $\|\wt V-\bSig_{\bzeta}\|_2
\leq\max _{k=1,\cdots,Kq}|\lambda_k\Big(\wh\L^{*}\Big)-\lambda_k\Big(n^{-1}\bzeta'_0\bzeta_0\Big)| + \|n^{-1}\bzeta_0\bzeta'_0-\bSig_{\bzeta}\|_2
\leq \|\wh\L^{*}-n^{-1}\bzeta'_0\bzeta_0\|_2+ \|n^{-1}\bzeta_0\bzeta'_0-\bSig_{\bzeta}\|_2=o_p(1)$ by Weyl’s Theorem and Condition (A1).
In addition, Condition (A1) shows $\|\bSig_{\bzeta}\|_2=O_p(1)$, which indicates $n^{-\frac{1}{2}}\|\wh\bzeta\|_2=O_p(1)$ because $\wt V=n^{-1}\wh\bzeta\wh\bzeta'$.
From Lemma 2 of \cite{asz010}, we have $\|\wt V^{-1}-\bSig^{-1}_{\bzeta}\|_2=o_p(1)$.
In addition, Condition (A1) also shows $\|\bSig^{-1}_{\bzeta}\|_2=O_p(1)$, which implies $\|\wt V^{-1}\|_2=O_p(1).$


Similar to the clues in the proof of the term $\Rmnum{1}_{b\zeta}$, we can get $\|\Rmnum{2}_{b\zeta}\|_2$ has the same order of $\|\Rmnum{1}_{b\zeta}\|_2$. It is easy to see that $\Rmnum{3}_{b\zeta}=o_p(\Rmnum{1}_{b\zeta})$ because $\|\wh\B-\B_0\|_F=o_p(\|\B_0\|_F)$.
Therefore, $\|\wh\bzeta_i-n^{-1}p^{-2}\tau_n^{-1}\wt V^{-1}
\Big(\sum_{j=1}^n\wh\bzeta_j\big(\sum_{l=1}^{n_j}\X'_{j}(t_{jl})\B_0\M^{*\prime}(t_{jl})\big)\K^{-1}_j\Big)\K^{-1}_i\big(
\sum_{l=1}^{n_i}\M^{*}(t_{il})\B'_0\X_i(t_{il})
\big)\|_2=\\O_p(n^{-\frac{1-e-v}{2}}+n^{\frac{e+v}{2}}p^{-\frac{1}{2}}+n^{\frac{e+v}{2}}N_0^{-1/2}).$

We then denote $H=n^{-1}\wt V^{-1}\wh\bzeta\bzeta'_0$ and get
\begin{equation}\label{eq:zeta_1}
\begin{split}
\wh\bzeta_i-H\bzeta_{i0}&=\frac{1}{np^2\tau_n}\wt V^{-1}
\Big(\sum_{j=1}^n\wh\bzeta_j\big(\sum_{l=1}^{n_j}\X'_{j}(t_{jl})\B_0\M^{*\prime}(t_{jl})\big)\K^{-1}_j\Big)\K^{-1}_i\big(
\sum_{l=1}^{n_i}\M^{*}(t_{il})\B'_0\X_i(t_{il})
\big)\\
&\ \ \ \ \ \ -\frac{1}{n}\wt V^{-1}\sum_{j=1}^n\wh\bzeta_j\bzeta'_{j0}\bzeta_{i0}
+O_p(n^{-\frac{1-e-v}{2}}+n^{\frac{e+v}{2}}p^{-\frac{1}{2}}+n^{\frac{e+v}{2}}N_0^{-1/2}).
\end{split}
\end{equation}
Substituting (\ref{eq:et}) into (\ref{eq:zeta_1}), we have 
\begin{equation}\label{eq:zeta_2}
\begin{split}
\wh\bzeta_i-H\bzeta_{i0}=
\frac{1}{n\tau_n}&\wt V^{-1}\sum_{j=1}^n\wh\bzeta_j\bzeta'_{j0}\Big(\sum_{l=1}^{n_j}\e(t_{jl})\M^{*\prime}(t_{jl})\Big)\K_j^{-1}\bTheta'_0\bzeta_{i0}\\
&+\frac{1}{np\tau_n}\wt V^{-1}\sum_{j=1}^n\wh\bzeta_j\left(\sum_{l=1}^{n_j}\u'_j(t_{jl})\B_0\M^{*\prime}(t_{jl})\K^{-1}_j\right)\bTheta'_0\bzeta_{i0}\\
&+\frac{1}{n\tau_n}\wt V^{-1}\sum_{j=1}^n\wh\bzeta_j\bzeta'_{j0}\bTheta_0\left(\sum_{l=1}^{n_i}\K^{-1}_i\M^{*}(t_{il})\e'(t_{il})\right)\bzeta_{i0}
\\
&+\frac{1}{n\tau_n}\wt V^{-1}\sum_{j=1}^n\wh\bzeta_j\bzeta'_{j0}\left(\sum_{l=1}^{n_j}\e(t_{jl})\M^{*\prime}(t_{jl})\K^{-1}_j\right)\left(\sum_{l=1}^{n_i}\K^{-1}_i\M^{*}(t_{il})\e'(t_{il})\right)\bzeta_{i0}\\
&+\frac{1}{np\tau_n}\wt V^{-1}\sum_{j=1}^n\wh\bzeta_j\left(\sum_{l=1}^{n_j}\u'_j(t_{jl})\B_0\M^{*\prime}(t_{jl})\K^{-1}_j\right)\left(\sum_{l=1}^{n_i}\K^{-1}_i\M^{*}(t_{il})\e'(t_{il})\right)\bzeta_{i0}\\
&+\frac{1}{np\tau_n}\wt V^{-1}\sum_{j=1}^n\wh\bzeta_j\bzeta'_{j0}\bTheta_0\left(\sum_{l=1}^{n_i}\K^{-1}_i\M^{*}(t_{il})\B'_0\u_i(t_{il})\right)\\
&+\frac{1}{np\tau_n}\wt V^{-1}\sum_{j=1}^n\wh\bzeta_j\bzeta'_{j0}\left(\sum_{l=1}^{n_j}\e(t_{jl})\M^{*\prime}(t_{jl})\K^{-1}_j\right)\left(\sum_{l=1}^{n_i}\K^{-1}_i\M^{*}(t_{il})\B'_0\u_i(t_{il})\right)\\
&+\frac{1}{np^2 \tau_n}\wt V^{-1}\sum_{j=1}^n\wh\bzeta_j\left(\sum_{l=1}^{n_j}\u'_j(t_{jl})\B_0\M^{*\prime}(t_{jl})\K^{-1}_j\right)\left(\sum_{l=1}^{n_i}\K^{-1}_i\M^{*}(t_{il})\B'_0\u_i(t_{il})\right)\\
&+O_p(n^{-\frac{1-e-v}{2}}+n^{\frac{e+v}{2}}p^{-\frac{1}{2}}+n^{\frac{e+v}{2}}N_0^{-1/2})\\
\triangleq\ \Rmnum{1}&+\Rmnum{2}+\Rmnum{3}+\Rmnum{4}+\Rmnum{5}+\Rmnum{6}+\Rmnum{7}+\Rmnum{8}+O_p(n^{-\frac{1-e-v}{2}}+n^{\frac{e+v}{2}}p^{-\frac{1}{2}}+n^{\frac{e+v}{2}}N_0^{-1/2}).
\end{split}
\end{equation}

For $\Rmnum{1}$, since for each $i$, we have
\begin{equation*}
\|\sum_{l=1}^{n_i}\e(t_{il})\M^{*\prime}(t_{il})\K_j^{-1}\bTheta_0\|_F\leq\left(\sum _{j=1}^{q}\sum _{k=1}^{K}\sum_{k'=1}^{K} \Big(\sum_{l=1}^{n_i}e_{jk}(t_{il})\bTheta'_{jk'0}\M(t_{il})\K^{-1}_i\Big)^2\right)^{\frac{1}{2}}=O_p(n^{-rv+v}).
\end{equation*}
Then by Cauchy's inequality, we have
$
\|\Rmnum{1}\|_2
\leq\frac{1}{n\tau_n}\|\wt V^{-1}\|_2 \cdot \|\wh\bzeta\|_F \cdot \|\bzeta_0\|_2  \cdot \|\bzeta_{i0}\|_2 \cdot O_p(n^{-rv+v}),
$
with $\|\bzeta_0\|_2=O_p(n^{\frac{1}{2}})$  by Condition (A1).
Therefore, $\|\Rmnum{1}\|_2=O_p(n^{-rv+\frac{e}{2}}).$
Similarly, we can get $\|\Rmnum{3}\|_2=O_p(n^{-rv+\frac{e}{2}}).$

For $\Rmnum{2}$, 
since for each $i$ we have  $\|\sum_{l=1}^{n_i}\u'(t_{il})\B_0\M^{*\prime}(t_{il})\K^{-1}_i\|_F=O_p(n^{-\frac{1-2v}{2}}p^{-\frac{1}{2}} )$ by Condition (A4), then we have
$\|\Rmnum{2}\|_2
\leq\frac{K}{np\tau_n}\|\wt V^{-1}\|_2 \cdot \|\wh\bzeta\|_F \cdot \|\bTheta_0\|_F \cdot \|\bzeta_{i0}\|_2 \cdot O_p(n^{-\frac{1-e-2v}{2}}p^{-\frac{1}{2}} )$
and $\|\bTheta_0\|_F=O(n^{\frac{v}{2}})$ by (I1). 
Thus $\|\Rmnum{2}\|_2=O_p(n^{\frac{e+v}{2}}p^{-\frac{1}{2}} ).$
Similarly, we have $\|\Rmnum{6}\|_2=O_p(n^{\frac{e+v}{2}}p^{-\frac{1}{2}})$.

Moreover, for each $i$, we have $\left\|\sum_{l=1}^{n_i}\e(t_{il})\M^{*\prime}(t_{il})\K^{-1}_i\right\|_2=O_p(n^{-rv+\frac{v}{2}})$.
Then we have
\begin{equation*}
\begin{split}
\|\Rmnum{4}\|_2\leq\frac{1}{n\tau_n}\|\wt V^{-1}\|_2 \cdot \|\wh\bzeta\|_F \cdot \|\bzeta_0\|_2  \cdot \|\bzeta_{i0}\|_2 \cdot O_p(n^{-2rv+v})=O_p(n^{-2rv+\frac{e}{2}}).
\end{split}
\end{equation*}
And $\|\Rmnum{5}\|_2\leq\frac{1}{np\tau_n}\|\wt V^{-1}\|_2 \cdot \|\wh\bzeta\|_F \cdot
\|\bzeta_{i0}\|_2 \cdot O_p(n^{-\frac{1-2v}{2}}p^{-\frac{1}{2}}) \cdot O_p(n^{-rv+\frac{v}{2}}) =O_p(n^{-rv+\frac{e+v}{2}}p^{-\frac{1}{2}}).$
Similarly, $\|\Rmnum{7}\|_2=O_p(n^{-rv+\frac{e+v}{2}}p^{-\frac{1}{2}})$. 

Similar to $\Rmnum{3}_b$, it follows 
\begin{equation*}
\begin{split}
\Rmnum{8}=
\frac{1}{np^2\tau_n}\wt V^{-1}&\sum_{j=1}^n\wh\bzeta_j\Ex\big[\big(\sum_{l=1}^{n_j}\u'_j(t_{jl})\B_0\M^{*\prime}(t_{jl})\K^{-1}_j\big)\big(\sum_{l=1}^{n_i}\K^{-1}_i\M^{*}(t_{il})\B'_0\u_i(t_{il})\big)\big]\\
&+\frac{1}{np^2\tau_n}\wt V^{-1}\sum_{j=1}^n\wh\bzeta_j
\left\{
\big(\sum_{l=1}^{n_j}\u'_j(t_{jl})\B_0\M^{*\prime}(t_{jl})\K^{-1}_j\big)\big(\sum_{l=1}^{n_i}\K^{-1}_i\M^{*}(t_{il})\B'_0\u_i(t_{il})\big)\right.\\
&-
\left.\Ex\big[\big(\sum_{l=1}^{n_j}\u'_j(t_{jl})\B_0\M^{*\prime}(t_{jl})\K^{-1}_j\big)\big(\sum_{l=1}^{n_i}\K^{-1}_i\M^{*}(t_{il})\B'_0\u_i(t_{il})\big)
\big]
\right\}.
\end{split}
\end{equation*}

By (\ref{eq:zeta_2}), 
we can conclude that
$\left(n^{-1}\sumi \|\wh\bzeta_i-\bzeta_{i0}\|^2_2\right)^{\frac{1}{2}}=O_p(\delta_n)=O_p(n^{-rv+\frac{e}{2}}
+n^{\frac{e+v}{2}}p^{-\frac{1}{2}}
+n^{-\frac{1-e-v}{2}}+n^{\frac{e+v}{2}}N_0^{-1/2}+n^{-1+\frac{e}{2}+v}+n^{\frac{e}{2}+v}p^{-1}).$
Then using the similar clues of Lemma A.2 in \cite{bai2003inferential}, 
we can get $\|\Rmnum{8}\|_2=O_p( n^{-1 + \frac{e}{2}+v}  + \delta_n n^{-\frac{1}{2} +\frac{e}{2}+v}
+ \delta_n n^{\frac{e}{2}+v}p^{-\frac{1}{2}} + n^{-\frac{1-e-2v}{2} }p^{-\frac{1}{2}} ). $
Therefore, $\|\wh\bzeta_i-H\bzeta_{i0}\|_2=O_p(n^{-\frac{1-e-v}{2}} + n^{\frac{e+v}{2}}p^{-\frac{1}{2}}  + n^{\frac{e+v}{2}}N_0^{-1/2}
+ n^{-rv+e} ).$

Now we show $H=\I_{Kq}+O_p(\delta_n n^{-\frac{1}{2}}).$
In fact, since $n^{-1}\wh\bzeta\bzeta'_0=n^{-1}(\wh\bzeta-H\bzeta_0+H\bzeta_0)\bzeta'_0= H\bSig_{\bzeta}+O_p(\delta_n n^{-\frac{1}{2}})$
and
$n^{-1}\wh\bzeta\bzeta'_0 H'=n^{-1}\wh\bzeta(\bzeta'_0 H'-\wh\bzeta'+\wh\bzeta')=\wt V+O_p(\delta_n n^{-\frac{1}{2}})$, we have $HH'=n^{-1}\wt V^{-1}\wh\bzeta\bzeta_0 H'+ O_p(\delta_n n^{-\frac{1}{2}})=
\wt V^{-1}\wt V + O_p(\delta_n n^{-\frac{1}{2}})
=\I_{Kq}  +  O_p(\delta_n n^{-\frac{1}{2}}) .$
This shows that $H'$ is an orthogonal matrix 
with the eigenvalues being either 1 or $-1$ up to the order of $O_p(\delta_n n^{-\frac{1}{2}})$.
By the definition of $H,$ we have
$H=n^{-1}\wt V^{-1}\wh\bzeta\bzeta'_0  +  O_p(\delta_n n^{-\frac{1}{2}})=\wt V^{-1} H\bSig_{\bzeta}$ then $\wt V H=H\bSig_{\bzeta} + O_p(\delta_n n^{-\frac{1}{2}}).$
It implies that $H$ is a diagonal matrix up to the order of $O_p(\delta_n n^{-\frac{1}{2}})$.
That is, $\|H-\I_{Kq}\|_F = O_p(\delta_n n^{-\frac{1}{2}}).$
Therefore, $\|\wh\bzeta_i-\bzeta_{i0}\|_2= \| \wh\bzeta_i-H\bzeta_{i0} \|_2 + \|H-\I_{Kq}\|_F \|\bzeta_{i0}\|_2 = O_p(n^{-\frac{1-e-v}{2}} + n^{-rv+\frac{e}{2}}  + n^{\frac{e+v}{2}}p^{-\frac{1}{2}} +n^{\frac{e+v}{2}}N_0^{-1/2} ).$

By (\ref{eq:zetab1}), (\ref{eq:zeta_2}) and the similar proof of Theorem 3.1 in \cite{bai2013statistical},
we have
\begin{equation}\label{eq:sup}
\begin{split}
\mathop{\sup}_i\|\wh\bzeta_i&-n^{-1}p^{-2}\tau_n^{-1}\wt V^{-1}
\Big(\sum_{j=1}^n\wh\bzeta_j\big(\sum_{l=1}^{n_j}\X'_{j}(t_{jl})\B_0\M^{*\prime}(t_{jl})\big)\K^{-1}_j\Big)\K^{-1}_i\big(
\sum_{l=1}^{n_i}\M^{*}(t_{il})\B'_0\X_i(t_{il})
\big)\|_2\\
&=O_p\{ (\log n)^{1/a_1} K^{1/2}\tau_n^{1/2}\|p^{-1/2}(\wh\B-\B) \|_F \}
+
O_p\{(\log n)^{1/a_2}\tau_n^{1/2}\|p^{-1/2}(\wh\B-\B) \|_F\}\\
&=O_p\{ (\log n)^{1/a_1} K^{1/2}\tau_n^{1/2}\|p^{-1/2}(\wh\B-\B) \|_F \}
.
\end{split}
\end{equation}
The uniform rate of $\|n^{-1}p^{-2}\tau_n^{-1}\wt V^{-1}
\Big(\sum_{j=1}^n\wh\bzeta_j\big(\sum_{l=1}^{n_j}\X'_{j}(t_{jl})\B_0\M^{*\prime}(t_{jl})\big)\K^{-1}_j\Big)\K^{-1}_i\big(
\sum_{l=1}^{n_i}\M^{*}(t_{il})\\\B'_0\X_i(t_{il})
\big)-\bzeta_{i0}\|_2$ is dominated by $\Rmnum{1}-\Rmnum{8}$.
Then following the similar proof of Theorem 3.1 in \cite{bai2013statistical}, we have $\mathop{\sup}_i\Rmnum{1}=\mathop{\sup}_i\Rmnum{3}=O_p((\log n)^{1/a_1}K^{1/2}\tau^{-r}_n),$
$\mathop{\sup}_i\Rmnum{2}=O_p((\log n)^{1/a_1}K^{1/2}\tau^{1/2}_np^{-1/2}),$
$\mathop{\sup}_i\Rmnum{4}=O_p((\log n)^{1/a_1}K^{1/2}\tau^{-2r}_n),$
$\mathop{\sup}_i\Rmnum{5}=O_p((\log n)^{1/a_1}K^{1/2}\tau^{1/2-r}_np^{-1/2}),$
$\mathop{\sup}_i\Rmnum{6}=O_p(n^{1/2\delta}K^{1/2}p^{-1/2}),$
$\mathop{\sup}_i\Rmnum{7}=O_p(n^{1/2\delta}K^{1/2}\tau^{-r}_np^{-
\frac{1}{2}}),$
and
$\mathop{\sup}_i\Rmnum{8}=O_p( K^{1/2}\tau^{1/2}_nn^{-1/2}+n^{1/2\delta}K^{1/2}\tau^{1/2}_np^{-1/2}).$

Finally, we have $\mathop{\sup}_i\|\wh\bzeta_i-\bzeta_{i0}\|_2=O_p\Big\{
 (\log n)^{1/a_1} K^{1/2}\big(\tau^{1/2}_n(n^{-1/2}+N^{-1/2}_0) + \tau^{-r}_n\big)
+n^{1/2\delta}K^{1/2}\tau^{1/2}_np^{-1/2}
\Big\}.$
This completes the proof of Theorem 2.
\\
\\

\textbf{Proof of Theorem 3:} The proof is similar to that of Theorem 2. By the definition of eigenvectors, we have $\wh\bTheta=n^{-1}p^{-2}\wt V^{-1}
\Big(\sumi\wh\bzeta_i\big(\sum_{l=1}^{n_i}\X'_i(t_{il})\wh\B\M^{*\prime}(t_{il})\big)\K^{-1}_i\Big)$, where $\wt V=n^{-1}\wh\bzeta\wh\bzeta'$ and
$\K_i=\sum_{l=1}^{n_i}\M^{*}(t_{il})\M^{*\prime}(t_{il})$
are the same as those in Theorem 2.
We note that
\begin{equation}\label{eq:theta1}
\begin{split}
\Big\|\wh\bTheta&-n^{-1}p^{-1}\wt V^{-1}
\Big(\sumi\wh\bzeta_i\big(\sum_{l=1}^{n_i}\X'_i(t_{il})\B_0\M^{*\prime}(t_{il})\big)\K^{-1}_i\Big)\Big\|_F\\
&\triangleq\Big\|
n^{-1}p^{-1}\wt V^{-1}
\Big(\sumi\wh\bzeta_i\big(\sum_{l=1}^{n_i}\X'_i(t_{il})(\wh\B-\B_0)\M^{*\prime}(t_{il})\big)\K^{-1}_i\Big)\Big\|_F\\
&=O_p((n^{-1/2}+p^{-1/2}+N^{-1/2}_0)\tau_nK^{1/2}),
\end{split}
\end{equation}
which is similar to the conclusion in (\ref{eq:zetab1}).
Then we show that
\begin{equation}\label{eq:theta2}
\begin{split}
\Big\|n^{-1}p^{-1}\wt V^{-1}
\Big(\sumi\wh\bzeta_i\big(\sum_{l=1}^{n_i}\X'_i(t_{il})&\B_0\M^{*\prime}(t_{il})\big)\K^{-1}_i\Big)-H\bTheta_0\Big\|_F\\
&=
\|n^{-1}\sumi\wh\bzeta_i\bzeta'_{i0}\sum_{l=1}^{n_i}\e(t_{il})\M^{*\prime}(t_{il})\K^{-1}_i\|_F\\
&\ \ \ + \|n^{-1}p^{-1}\wt V^{-1}\sumi\wh\bzeta_i\sum_{l=1}^{n_i}\u'_i(t_{il})\B_0\M^{*\prime}(t_{il})\K^{-1}_i\|_F,
\end{split}
\end{equation}
where $H=n^{-1}\wt V^{-1}\wh\bzeta\wh\bzeta'$ is the same as that in Theorem 2.
The first term in (\ref{eq:theta2}) is $O_p(K^{1/2}\tau^{-r+1/2}_n)$ and the second term is $O_p(K^{1/2}\tau_np^{-1/2})$,
which are similar to $\Rmnum{1}$ and $\Rmnum{2}$ in (\ref{eq:zeta_2}). We have shown that $\|H-\I_{Kq}\|_F$ is  sufficiently small, so we have $\|\wh\bTheta-\bTheta_0\|_F=O_p((n^{-1/2}+p^{-1/2}+N^{-1/2}_0)K^{1/2}\tau_n+K^{1/2}\tau^{-r+1/2}_n)$.
And $\|\wh\bTheta_j-\bTheta_{j0}\|_F$ is of the same order as $\|\wh\bTheta-\bTheta_0\|_F$ for each $j=1,\cdots,q$ because $q$ is fixed.
Finally we have $\|\wh\PPhi_j(t)-\PPhi_{j0}(t)\|_F=O_p((n^{-1/2}+p^{-1/2}+N^{-1/2}_0)K^{1/2}\tau^{1/2}_n+K^{1/2}\tau^{-r}_n)$ by noting $\tau_n\int\M(t)\M'(t)dt=\I_{\tau_n}.$ Thus we complete the proof.
\\
\\

\textbf{Proof of Theorem 4:}
By the proof procedure in Theorem 1, for each $j$, the order of $\wh\b_{j}-\b_{j0}$ is dominated by $\Rmnum{1}_b$, thus we have
\begin{equation}\label{b:an}
\begin{split}
\sqrt{n}(\wh\b_{j}-\b_{j0})=&\sqrt{n}(\wh\b_{j}-\wc\H\b_{j0})+o_p(1)\\
=&\wc V^{-1} \frac{1}{p}\wh\B'\B_0\frac{1}{\sqrt{n}}\sumi\sum_{l=1}^{n_i}\frac{1}{n_i}\PPhi_0(t_{il})
\bzeta_{i0}u_{ij}(t_{il})+o_p(1),\\
=&\wc V^{-1} \frac{1}{p}\wh\B'\B_0\frac{1}{\sqrt{n}}\sumi\int\PPhi_0(t)
\bzeta_{i0}u_{ij}(t)dt+o_p(1).
\end{split}
\end{equation}
based on (\ref{eq:wcb1}) and numerical integration approximation.
We have shown that $\wc V=\LLambda_{\bzeta}+o_p(1)$ and $\wc V^{-1}=\LLambda_{\bzeta}^{-1}+o_p(1)$ in the proof of Theorem 1. Further, $p^{-1}\wh\B'\B_0=p^{-1}(\wh\B'-\B_0)\B_0+p^{-1}\B'_0\B_0=\I_q+O_p(n^{-\frac{1}{2}}+p^{-\frac{1}{2}}).$
Finally, we have
\begin{equation*}
\sqrt{n}(\wh\b_{j}-\b_{j0})=\LLambda_{\bzeta}^{-1}\frac{1}{\sqrt{n}}\sumi\int\PPhi_0(t)
\bzeta_{i0}u_{ij}(t)dt+o_p(1)
\end{equation*}
and the desired limiting distribution follows from the central limit theorem.


\bibliographystyle{apalike}
\bibliography{myref}